\documentclass[reprint,twocolumn,superscriptaddress,secnumarabic,amssymb, nobibnotes, aps, prb]{revtex4-2}

\usepackage{amsmath}    
\usepackage{graphicx}    
\usepackage{verbatim}    
\usepackage{xcolor}           
\usepackage{subfigure}   
\usepackage{hyperref}    
\usepackage{mathrsfs}
\usepackage{braket}
\usepackage{physics}
\usepackage{graphicx}    
\usepackage{xspace}
\usepackage{bbold}

\newcommand{\bR}{\mathbf{R}}

\newcommand{\bs}{\mathbf{s}}

\newcommand{\bk}{\mathbf{k}}
\newcommand{\bq}{\mathbf{q}}

\newcommand{\bv}{\mathbf{v}}
\newcommand{\jh}{J_{\rm H}}

\def\ruo{RuO$_{2}$\xspace}

\begin{document}

\title{Incipient magnetic instability in RuO$_{\mathbf{2}}$ with random phase approximation}%

\author{Diana Csontosov\'a}
\affiliation{Department of Condensed Matter Physics, Faculty of
  Science, Masaryk University, Kotl\'a\v{r}sk\'a 2, 611 37 Brno,
  Czechia}
\author{Kyo-Hoon Ahn}
\affiliation{Institute of Physics, Czech Academy of Sciences, Cukrovarnick\'{a} 10, 162 00 Praha 6, Czechia}
\author{Jan Kune\v{s} }
\affiliation{Department of Condensed Matter Physics, Faculty of
  Science, Masaryk University, Kotl\'a\v{r}sk\'a 2, 611 37 Brno,
  Czechia}

\date{\today}

\begin{abstract}
We study the instability in \ruo using the Hartree–Fock approximation followed by the random phase approximation. We employ a three-orbital Hubbard model without spin–orbit coupling. An analysis of the eigenvalues and eigenvectors of the static susceptibility in the non-magnetic phase for various local interaction parameters $U$, $\jh$, and hole doping $n$ shows that the spin susceptibility is the dominant response channel.
In the stoichiometric system without spin-orbit coupling, commensurate altermagnetic order is identified as the leading instability at sufficiently low temperatures, whereas at higher temperatures or finite hole doping, incommensurate wave vectors emerge. To elucidate the origin of the magnetic instability, we analyze the band spitting by the staggered Weiss field and discuss the qualitative difference between altermagnets and antiferromagnets.
\end{abstract}

\maketitle


\section{Introduction}

\ruo, a metallic compound with the rutile structure, has attracted significant attention in recent years. 
Long considered a Pauli paramagnet~\cite{Guthrie1931, Ryden1970, CORDFUNKE1989, Lin2004}, an antiferromagnetic (AFM) order with small Ru moments of $0.05\,\mu_{\mathrm{B}}$ was reported by Berlijn~\textit{et~al.}~\cite{Berlijn2017} in 2017 and subsequently supported by resonant x-ray scattering
experiments~\cite{Zhu2019}. While the early density functional theory (DFT) studies based on local density approximation~\cite{MEHTOUGUI2012331} 
led to non-magnetic solution, inclusion of electronic correlations within the Ru $4d$ shell by means of DFT+U~\cite{Berlijn2017, SmejkalAHE2020, Liag2022, Smolyanyuk2024}, Hartree–Fock (HF) approximation or dynamical mean-field theory~\cite{Ahn2019}
led to a staggered magnetic order, albeit with substantially larger moments and only for sufficiently strong in-site interaction.
The origin of the magnetic order was attributed to Fermi-surface instability associated with symmetry-protected nodal lines~\cite{Sun2017, Jovic2018} located near the Fermi level~\cite{Berlijn2017, Ahn2019}.

Following the introduction of altermagnetism (AM)~\cite{Smejkal2022}, \ruo became a prototype of an altermagnet, exhibiting large spin splitting of the electronic bands of up to $\approx 1\,\mathrm{eV}$~\cite{Ahn2019, SmejkalAHE2020, Smejkal2023}. The existence of AFM phase was also supported by the observation of a large anomalous Hall effect (AHE) above $50\,\mathrm{T}$ in \ruo epitaxial thin films~\cite{Feng2022, Tschirner2023}.

Despite the  
evidence supporting magnetic order, the magnetic state of \ruo remained controversial. No clear thermodynamic anomaly indicating phase transition has been observed in the heat-capacity~\cite{CORDFUNKE1989}, and much of the experimental evidence supporting the magnetic order was obtained in thin films rather than in bulk single-crystal samples~\cite{HUSSAIN2025, li2026}. Angle-resolved photoemission spectroscopy (ARPES) 
did not lead to a unambiguous conclusions~\cite{Fedchenko2024, lin2024, Liu2024, Osumi2026}. 
Sensitivity of the sample stoichiometry and defects was proposed to explain the controversy
 shifting the attention to the role of doping 
 ~\cite{Wang2023, Smolyanyuk2024}, as well as toward the optimization of the growth processes to obtain high-purity samples.
 The $\mu$SR measurements~\cite{Kessler2024, Hiraishi2024} as well as revised neutron diffraction experiment~\cite{Kessler2024} concluded negligibly small magnetic moments on the Ru atoms in both single-crystal samples and thin films
 \footnote{The apparent contradiction with earlier neutron and resonant X-ray scattering results has been attributed to extrinsic effects, such as multiple scattering or sample-dependent structural defects; see Refs.~\cite{Kessler2024, Hiraishi2024} for a detailed discussion.}. These results were followed by transport~\cite{Peng2025} and optical~\cite{Wenzel2025} experiments, as well as quantum-oscillation measurements of the thermodynamic properties of bulk \ruo~\cite{Wu2025}, all of which indicate that the non-magnetic phase provides a better description of the experimental observations. The AHE measured in Cr-doped \ruo~\cite{Wang2023} was argued to arise from weak ferromagnetism due to the Cr dopants~\cite{Smolyanyuk2025}. 
 Despite the mounting  evidence of non-magnetic nature of \ruo, 
 the interest in its magnetic properties has not disappeared. Surface magnetism associated with the local symmetry breaking was proposed in a DFT study of Ho~\textit{et al.}~\cite{Ho2025}, even though the bulk remained non-magnetic. Magnetic ordering was also reported in epitaxial \ruo/TiO$_2$ heterostructures~\cite{Jeong2025} and in highly strained epitaxial $(110)$ \ruo thin films grown on $(110)$ TiO$_2$ substrates~\cite{Occhialini2022}.

In this work, we analyze the three-orbital Hubbard model of \ruo spanned by the Ru $t_{2g}$ by means of HF and random phase approximation (RPA) approaches. 
First, we study the static susceptibility $\hat{\chi}(\bq, \omega = 0)$ in the non-magnetic (NM) phase. This allows us to identify the instability that characterizes the possible ordering in the material. To understand the microscopic origin of this instability, we then analyze the NM and AM Fermi surfaces (FS), and the $k$-resolved band contribution to the condensation energy $\eta(\bk)$. Building on previous theoretical studies of the \ruo band structure~\cite{Berlijn2017, Jovic2018, Ahn2019}, we identify hot spots that could potentially lead to the emergence of AM order. We further investigate the effects of doping and an applied staggered potential, which lifts the fourfold symmetry, on the tendency of the material to magnetically order.

\begin{figure}
    \centering
    \includegraphics[width=0.9\linewidth]{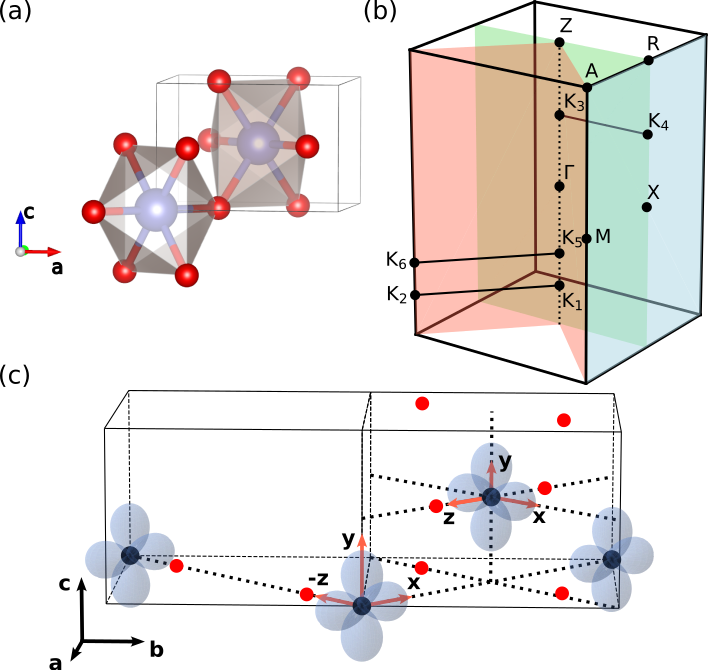}
    \caption{(a) Crystal structure of \ruo. Red spheres represent oxygen ions, and blue spheres represent ruthenium ions. The visualization was created using VESTA3~\cite{vesta}. (b) First Brillouin zone with highlighted planes where $\eta(\bk)$ is shown and paths along which the band structures are analyzed. (c) Schematic illustration of the direct hopping path between $d_{x^2 - y^2}$ orbitals on the same sublattice. Orange arrows indicate the local coordination, while $\mathbf{a}$, $\mathbf{b}$, and $\mathbf{c}$ denote the global axes. Red and black dots represent oxygen and ruthenium ions, respectively.}
    \label{fig:ruo_lattice_bz}
\end{figure}


\section{Model and computational methods}

\ruo crystallizes in the rutile structure with space group $P4_2/mnm$ shown in Fig.~\ref{fig:ruo_lattice_bz}~(a). The DFT band structure, calculated using the WIEN2K package~\cite{wien2k}, was represented on the basis of Ru-centered
Wannier orbitals with approximate $t_{2g}$ symmetry~\cite{KUNES20101888, MOSTOFI20142309}. 

The three-orbital Hubbard model with a two-atom unit cell is given by
\begin{align}
H &= H_0 + H_{\rm int},  \\[4pt]
H_0 &=
\begin{aligned}[t]
&\sum_{\mathbf{R}\mathbf{R}'} \sum_{\mathbf{s}\mathbf{s}'} \sum_{\ell m} \sum_{\sigma}
t^{\ell m}_{\mathbf{R}\mathbf{R}', \mathbf{s}\mathbf{s}'}
c^\dagger_{\mathbf{R}\mathbf{s}\ell \sigma}
c_{\mathbf{R}'\mathbf{s}'m\sigma} \\
& - \mu \sum_{\mathbf{R}\mathbf{s}} \sum_{\ell\sigma}
c^\dagger_{\mathbf{R}\mathbf{s}\ell\sigma} c_{\mathbf{R}\mathbf{s}\ell\sigma}
\end{aligned} \notag \\[3pt]
H_{\rm int} &= \sum_{\mathbf{R}\mathbf{s}} H^{\rm int}_{\mathbf{R}\mathbf{s}}. \notag 
\label{eq:H_total} 
\end{align}
Here, $\mathbf{R}$ and $\mathbf{R}'$ denote the positions of the unit cells, and $\mathbf{s}, \mathbf{s}' \in \{\mathbf{s}_1, \mathbf{s}_2\}$ are the atomic positions within the unit cell. In our case, $\mathbf{s}_1 = (0, 0, 0)$ and $\mathbf{s}_2 = (0.5, 0.5, 0.5)$ (see Fig.~\ref{fig:ruo_lattice_bz}~(a)). The indices $\ell$ and $m$ run over orbital flavors, and $\sigma$ denotes the spin index. The operators $c^\dagger$ and $c$ are fermionic creation and annihilation operators, respectively, $t^{\ell m}_{\mathbf{R}\mathbf{R}', \mathbf{s}\mathbf{s}'}$ is the hopping amplitude, and $\mu$ denotes the chemical potential. The spin-orbit coupling (SOC) was neglected. The local interaction is described by the Slater–Kanamori Hamiltonian,
\begin{equation}
\label{eq:ruo2_Hint}
\begin{aligned}
H^{\rm int}_{\mathbf{R}\mathbf{s}} &= U \sum_{\ell} n_{\ell\uparrow} n_{\ell\downarrow} \\
&\quad + \sum_{\ell>m,\sigma\sigma'}
\left( U - 2J_{\mathrm{H}} - J_{\mathrm{H}}\delta_{\sigma\sigma'} \right)
n_{\ell\sigma} n_{m\sigma'} \\
&\quad + J_{\mathrm{H}} \sum_{\ell\neq m,\sigma}
\Bigl(
c_{\ell -\sigma}^{\dagger} c_{m \sigma}^{\dagger}
c_{\ell\sigma}^{\phantom\dagger} c_{m -\sigma}^{\phantom\dagger} \\
&\quad - c_{\ell-\sigma}^{\dagger} c_{\ell \sigma}^{\dagger}
c_{m\sigma}^{\phantom\dagger} c_{m -\sigma}^{\phantom\dagger}
\Bigr)
\end{aligned}
\end{equation}
where $n_{\ell\sigma}$ is the particle-number operator $c_{\ell \sigma}^{\dagger} c_{\ell \sigma}$. For clarity, we omit the $\mathbf{R}$ and $\mathbf{s}$ indices in the definition of the local interaction.

Since \ruo is considered a weak-coupling metal and it has been shown that dynamical correlations do not qualitatively modify the band structure~\cite{Ahn2019}, we employ the Hartree–Fock (HF) approximation, in which the self-energy $\Sigma_{\mathrm{HF}}$ is treated as static. One-particle quantities, such as the occupation matrix and the chemical potential $\mu$, are obtained from self-consistent calculations. The Fermi velocity was inspected by the FermiSurfer~\cite{Kawamura2018FermiSurferFV} viewer.

In order to calculate the static susceptibility $\hat{\chi}(\bq, \omega=0)$ and the magnon spectra, we employ the RPA approximation on the real-frequency axis,
\begin{equation}
\hat{\chi}(\bq, \omega) = \hat{\chi}_0(\bq, \omega)\left[\mathbb{1} + \hat{U}\hat{\chi}_0(\bq, \omega)\right]^{-1}.
\label{eq:RPA}
\end{equation}
Here, $\hat{U}$ denotes the interaction matrix that has, as well as other quantities in Eq~\eqref{eq:RPA}, size $72\times 72$. The bubble $\chi_0$ on the imaginary-frequency axis is given by the product of one-particle Green's functions
\begin{equation}
    G_{\ell \sigma_1 s, n \sigma_3 s'}(\bk, i\nu) = \int_0^\beta \mathrm{d}\tau e^{i\nu \tau} \langle c_{\bk s\ell \sigma_1}(\tau) c^\dagger_{\bk s'n\sigma_3}(0) \rangle
\end{equation}
as
\begin{equation}
    \begin{aligned}
        \chi^{\ell \sigma_1 m \sigma_2 s, n \sigma_3 p \sigma_4 s^{\prime}}_0(\bq, i\nu, i\nu', i \omega) =
      - \frac{1}{N_k}\sum_{\bk}
      & G_{\ell \sigma_1 s, n \sigma_3 s^{\prime}}(\bk, i\nu) \\
      \times 
      G_{p \sigma_4 s^{\prime}, m \sigma_2 s}(\bk + \bq, i\nu' + i\omega)\delta_{\nu \nu'}.
    \end{aligned}
\end{equation}
\begin{figure*}
    \centering
    \includegraphics[width=1.\linewidth]{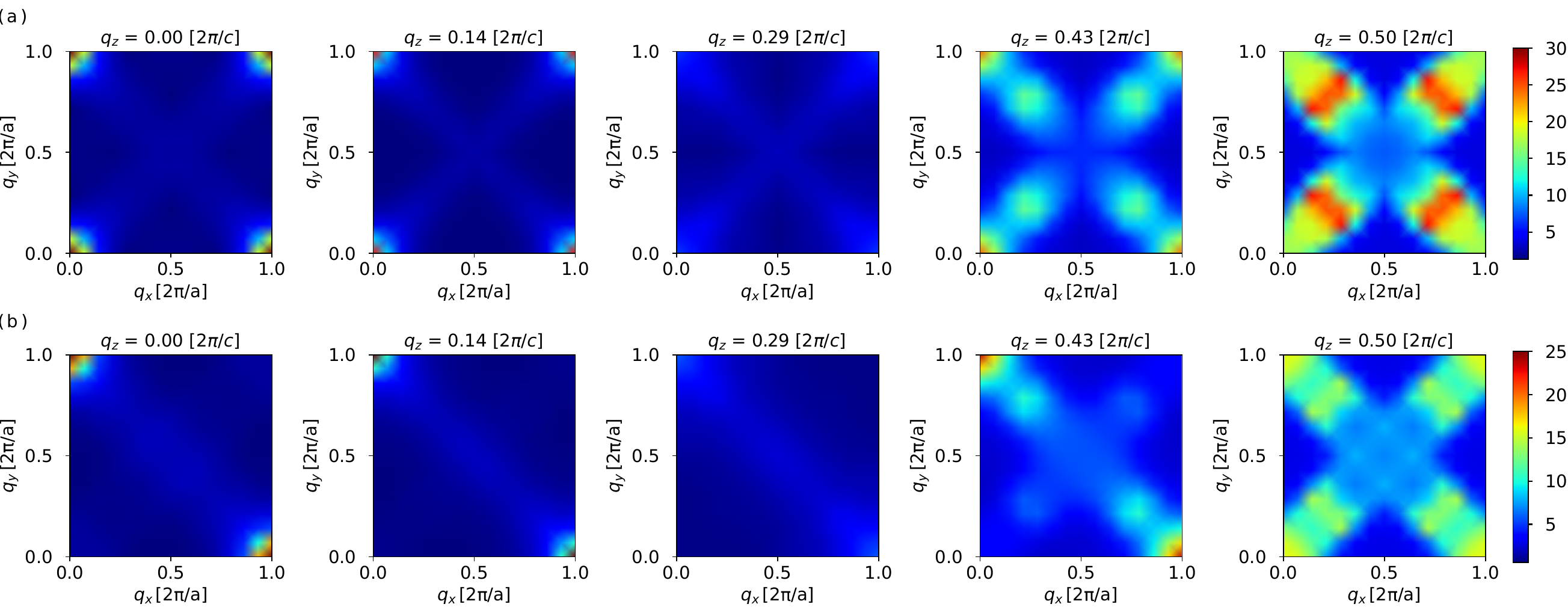}
    \caption{(a) Leading eigenvalue $\Lambda(\bq)$ of the static susceptibility $\chi(\bq)$ 
    ($U = 1.24\,\mathrm{eV}$, $J_{\mathrm{H}} = 0.3\,\mathrm{eV}$, $n = 0$ and $T = 232\,\mathrm{K}$) and (b) unfolded spin susceptibility $\tilde{\chi}(\bq)$ shown for various cuts in $q_z$ within the first Brillouin zone.}
    \label{fig:chi_stat_unf_ruo2}
\end{figure*}
\begin{figure}
    \centering
    \includegraphics[width=1.\linewidth]{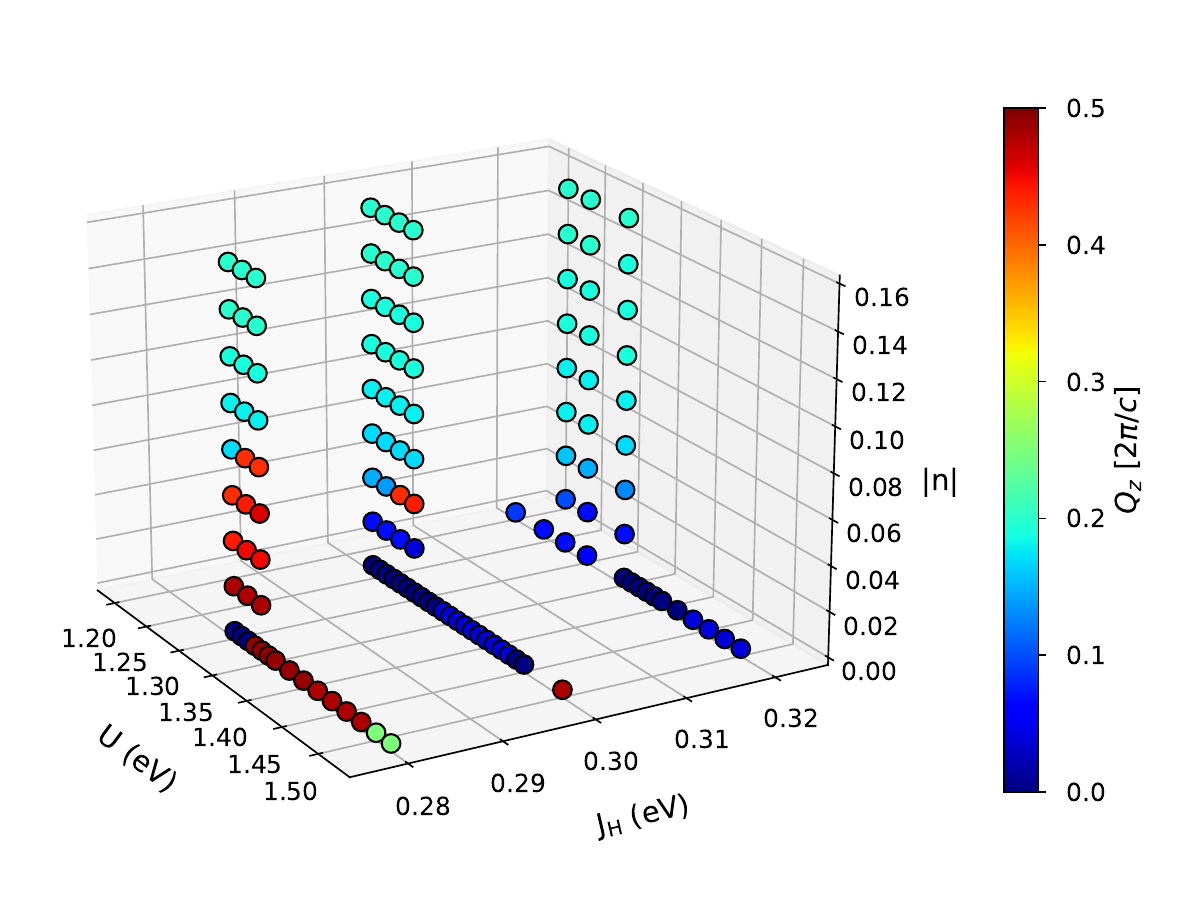}
    \caption{The position of the instability along the $q_z$ axis in $\mathbf{Q} = (0, 0, Q_z)$ as a function of the interaction parameters $U$, $\jh$, and the absolute value of the hole doping $|n|$, expressed in electrons per atomic site.}
    \label{fig:phase_diag_ruo2}
\end{figure}
%
The indices $n$ and $p$ label the orbitals. 
Since the one-particle Hamiltonian is spin diagonal, the Green's function is diagonal in spin indices $G_{\ell \sigma_1 s, n \sigma_3 s'}\sim\delta_{\sigma_1\sigma_3}$.
The frequencies $i\nu$ and $i\omega$ denote fermionic and bosonic Matsubara frequencies, respectively, $\tau$ is the imaginary time, $s, s'$ run over site indices, and $N_k$ denotes the number of $k$ points. The operator $c_{\bk\ell \sigma s}$ is related to $c_{\mathbf{R}\mathbf{s}\ell}$ by a Fourier transformation $c_{\bk s \sigma \ell} = \frac{1}{\sqrt{N}} \sum_{\mathbf{R}} e^{i\bk\mathrm{R}} c_{\mathbf{R}\mathbf{s}\ell}$.

Within the HF approximation, the bubble defined on the imaginary-frequency axis can be transformed to the real-frequency axis by relatively straightforward analytical calculation, shown in the Supplementary material~\cite{SupplementRuO2}, yielding
\begin{equation}
\label{eq:bubble}
\begin{aligned}
\chi_0&^{\ell \sigma_1 m \sigma_2 s, n \sigma_3 p \sigma_4 s^{\prime}}(\bq, \omega) = \frac{1}{N_k} \sum_{ij} \sum_{k} w^i_{\ell \sigma_1 s, n \sigma_3 s^{\prime}}(\bk)
\times \\
&w^j_{p \sigma_4 s^{\prime}, m \sigma_2 s}(\bk+\bq) 
\frac{f_{\mathrm{FD}}(E_i(\bk)) - f_{\mathrm{FD}}(E_j(\bk+\bq))}{\omega + i\varepsilon+ E_i(\bk) - E_j(\bk+\bq)},
\end{aligned}
\end{equation}
where $\hat{w}_i = \ket{i}\bra{i}$ and $\hat{w}_j = \ket{j}\bra{j}$ are the outer products of the eigenvectors of $\Tilde{H}_k = H_k + \Sigma_{\mathrm{HF}}$ corresponding to eigenvalues $E_i$ and $E_j$, respectively, and $\varepsilon$ is infinitesimal. Here, $f_{\mathrm{FD}}$ denotes the Fermi–Dirac distribution. Finally, the resulting rank-four tensor is reshaped into a matrix form suitable for use in Eq.~\eqref{eq:RPA}.

The interaction matrix $\hat{U}$ has the following structure:
\begin{subequations}
    \begin{align}
        (U_{\uparrow\downarrow})_{\ell\ell,\ell\ell} &= U, \\
        (U_{\uparrow\downarrow})_{\ell\ell,mm} &= U', \\
        (U_{\uparrow\downarrow})_{\ell m,\ell m} &= \jh, \\
        (U_{\uparrow\downarrow})_{\ell m,m\ell} &= \jh,
    \end{align}
\end{subequations}
\begin{subequations}
    \begin{align}
        (U_{\uparrow\uparrow})_{\ell\ell,mm} &= U'-\jh, \\
        (U_{\uparrow\uparrow})_{\ell m,\ell m} &= \jh - U',
    \end{align}
\end{subequations}
\begin{subequations}
    \begin{align}
        (U_{\overline{\uparrow\downarrow}})_{\ell\ell,\ell\ell} &= -U, \\
        (U_{\overline{\uparrow\downarrow}})_{\ell\ell,mm} &= -\jh, \\
        (U_{\overline{\uparrow\downarrow}})_{\ell m,\ell m} &= -U', \\
        (U_{\overline{\uparrow\downarrow}})_{\ell m,m\ell} &= -\jh,
    \end{align}
\end{subequations}
where $U' = U - 2\jh$, and $\ell,m$ denote orbital indices. The lower indices $\uparrow\downarrow$, $\uparrow\uparrow$, and $\overline{\uparrow\downarrow}$ are abbreviations for $\{\uparrow\uparrow,\downarrow\downarrow\}$, $\{\uparrow\uparrow,\uparrow\uparrow\}$, and $\{\uparrow\downarrow,\uparrow\downarrow\}$, respectively. For both spin and orbital indices, those preceding the comma correspond to the rows of the $\hat{U}$ matrix, while those following the comma correspond to the columns. We note that the matrix elements are symmetric under the exchange of spins $\sigma \rightarrow -\sigma$.

Finally, we perform the unfolding of the susceptibility $\chi(\bq, \omega)$ to the one-atomic unit cell using the formula
%
\begin{equation}
    \begin{aligned}
        \tilde{\chi}_{\ell \sigma_1 m \sigma_2, n \sigma_3 p \sigma_4}(\bq, \omega) &= \frac{1}{2} \sum_{\mathbf{s}, \mathbf{s}'} e^{-i\bq(\mathbf{s}-\mathbf{s}')} \\
        &\times \chi_{\ell \sigma_1 m \sigma_2 s,n \sigma_3 p \sigma_4 s'}(\bq, \omega).
    \end{aligned}
\end{equation}
%

\begin{figure}
    \centering
    \includegraphics[width=1.\linewidth]{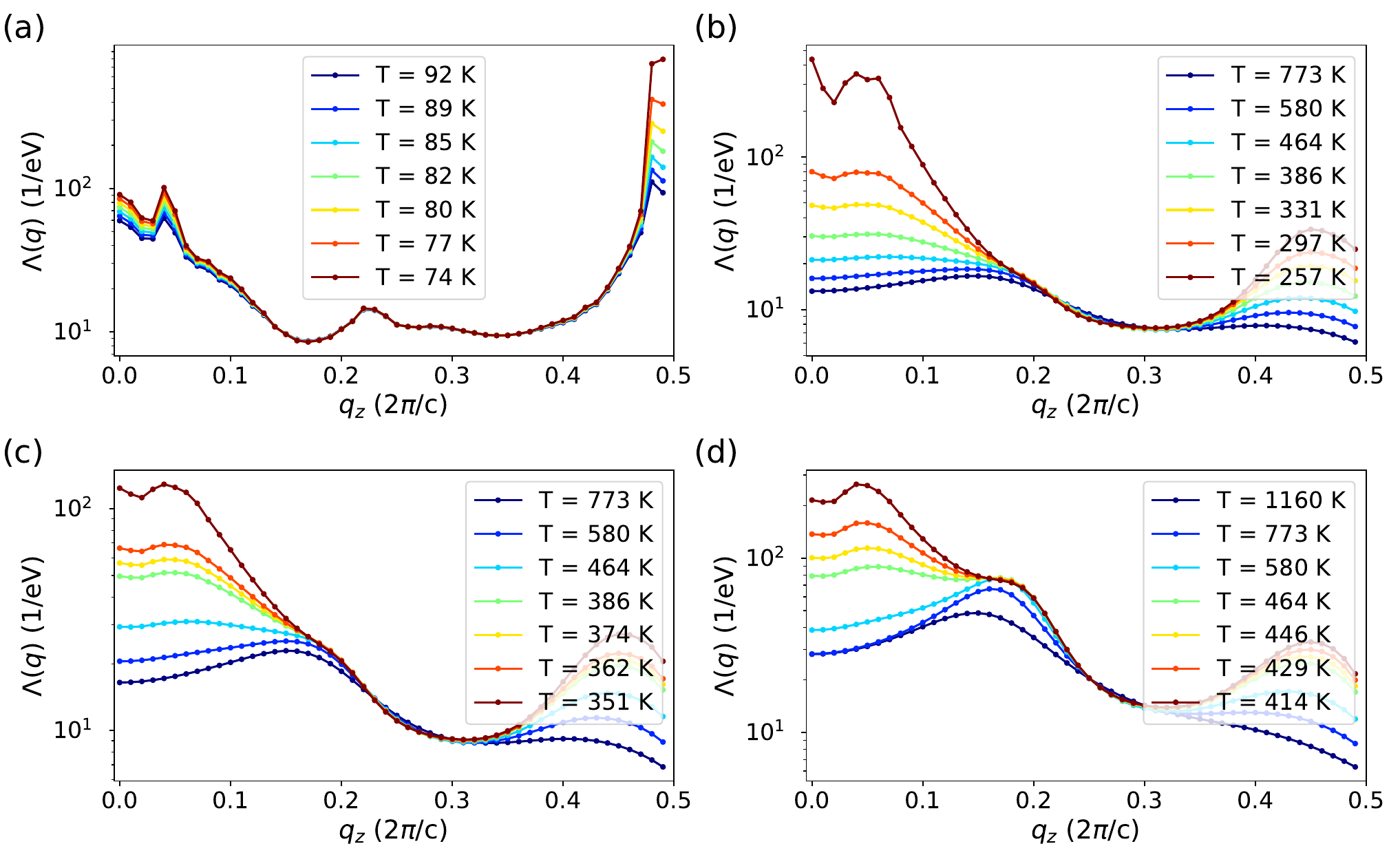}
    \caption{The temperature dependence of the largest eigenvalue $\Lambda(\bq)$ of the static susceptibility $\chi(\bq)$ for $\bq = (0, 0, q_z)$ in the NM phase. Different interaction parameters are compared: (a) $U = 1.35\,\mathrm{eV}$, $\jh = 0.28\,\mathrm{eV}$; (b) $U = 1.27\,\mathrm{eV}$, $\jh = 0.3\,\mathrm{eV}$; (c) $U = 1.3\,\mathrm{eV}$, $\jh = 0.3\,\mathrm{eV}$; (d) $U = 1.35\,\mathrm{eV}$, $\jh = 0.3\,\mathrm{eV}$.}
    \label{fig:eigvals_temp_dep_ruo2}
\end{figure}

\subsection{Analysis of instability of the non-magnetic phase}
An instability towards long-range order is signaled by a divergence of the static susceptibility $\chi(\bq)\equiv \hat{\chi}(\bq, 0)$ in the non-magnetic  phase. 
Its leading, largest, eigenvalues and corresponding eigenvectors provide information about the type of the order parameter although they do not
characterize it completely in case of multicomponent order parameters. This is the case when higher than quadratic terms in the Ginzburg-Landau functional are needed
to determine the order parameter, while the susceptibility is by definition the second order expansion.

Naively we could identify the largest eigenvalue $\Lambda(\bq)$ of $\chi(\bq)$, find the corresponding eigenvector and link it to electronic structure features such as Fermi surface nesting. The problem is that only $\chi_0(\bq)$ can be expressed as Brillouin zone (BZ) sum, while $\chi(\bq)$ given by Eq.~\eqref{eq:RPA} 
is not directly linked to the electronic structure. The spectra of $\chi_0(\bq)$ and $\chi(\bq)$ do not match, as the corresponding matrices do not commute
In order to make meaningful comparison of $\chi_0(\bq)$ and $\chi(\bq)$ we decided to work with a fixed external field/eigenvector.
\begin{figure}[t]
    \centering
    \includegraphics[width=0.75\linewidth]{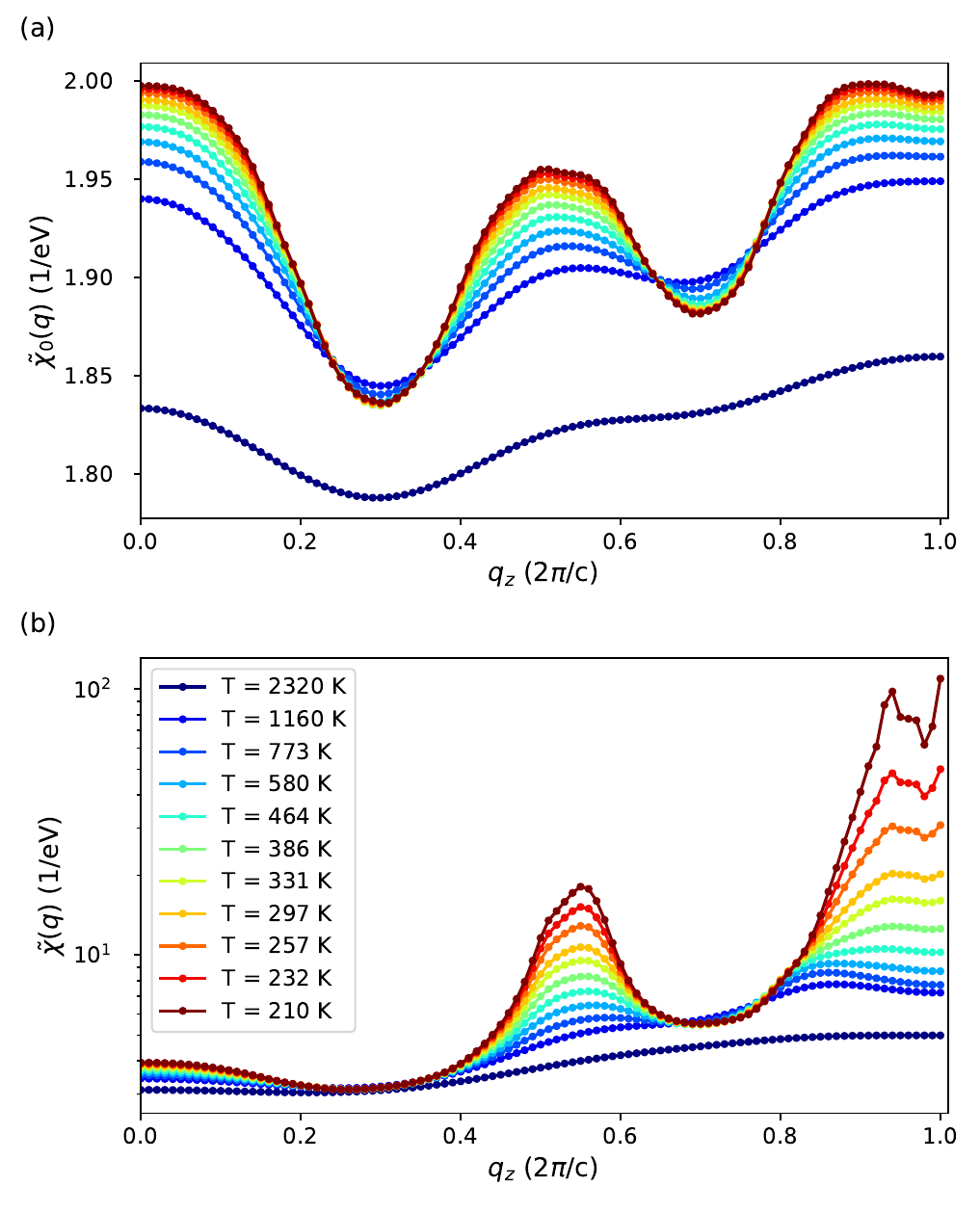}
    \caption{Comparison of (a) the static spin bubble $\tilde{\chi}_0(\bq)$ and (b) the static spin susceptibility $\tilde{\chi}(\bq)$ after the unfolding for $\bq = (0, 0, q_z)$ in the NM phase for different temperatures. The corresponding temperatures interaction parameters are $U = 1.24\,\mathrm{eV}$, $\jh = 0.3\,\mathrm{eV}$ and no doping.}
    \label{fig:bubble_vs_chi_ruo2}
\end{figure}

This poses another question. A casual examination of the leading eigenvectors suggests that they correspond to the spin susceptibility, i.e., a response to a Zeeman splitting 
\begin{equation}
\mathbf{Z}(\bq)=\sum_\bR\sum_{l=1}^3\sum_{\bs=\bs_1,\bs_2} \boldsymbol{\sigma}^{\phantom\dagger}_{\alpha\beta}e^{i\mathbf{q}(\bR+\bs)}c^\dagger_{\bR\bs l\alpha} c^{\phantom\dagger}_{\bR\bs l\beta},
\end{equation}
where $\boldsymbol{\sigma}$ denotes the Pauli matrix. However, since the orbitals are not equivalent, e.g., the occupation is rather different, the eigenvectors of $\chi(\bq)$ are not exactly $\mathbf{Z}(\bq)$.
How do we quantify how much a given eigenvector coincides with a response to a given external field? We propose to use an overlap 
\begin{equation}
\label{eq:measure}
    \mathcal{O}[\bv] = \frac{\{\mathbf{v}, \mathbf{Z}\}}{\sqrt{\{\mathbf{v}, \mathbf{v}\}\{\mathbf{Z}, \mathbf{Z}\}}},
\end{equation}
with a scalar product
\begin{equation}
\{\mathbf{v}, \mathbf{Z}\} \equiv \sum_{ijkl}\overline{\mathbf{v}^{ij}} \chi^{ij,kl} \mathbf{Z}^{kl}.
\end{equation}
Here, $\bv$ is the analyzed eigenvector, $\mathbf{Z}$ is reference, e.g., Zeeman, field and $\chi^{ij,kl}$ is the susceptibility matrix, all taken
in a given $\bq$-point in BZ (not shown for simplicity). One way to read this formula is that instead of comparing $\bv$ with $\mathbf{Z}$ directly,
we compare $\bv$ with the response of the system generated by $\mathbf{Z}$.

\section{Results}
\subsection{Instability of the non-magnetic phase}

We start with examining the character of the leading instability in \ruo for various interaction parameters $U$ and $\jh$, as well as for different levels of hole doping $n$. We find the largest (leading) eigenvalue $\Lambda(\bq)$ of the static susceptibility $\chi(\bq)$ in non-magnetic phase 
and analyze corresponding leading eigenvectors.
Throughout the studied parameter range 
the leading eigenvectors fall into the spin-triplet sector~\footnote{Thus form a three dimensional subspace.}.
In Fig.~\ref{fig:chi_stat_unf_ruo2}(a) we show 
the typical dependence $\Lambda(\bq)$ with maximum at $(0, 0, Q_z)$
observed throughout the studied parameters. 
Here, $Q_z=0$, but in general it varies across the parameter space, see Fig.~\ref{fig:phase_diag_ruo2}. In particular,
it varies with temperature as shown in Fig.~\ref{fig:eigvals_temp_dep_ruo2}.


For the undoped system we observe a general trend of $Q_z$ moving towards 0 or $\pi$ at sufficiently low temperature. These temperatures may be below
the generally overestimated RPA transition temperatures. To identify the 'leading instability' in such cases we perform constrained NM calculations and look at the lowest inverse $\Lambda^{-1}(q)$, which is then negative, see Supplementary Figure~1~\cite{SupplementRuO2}.
With hole doping $n$, the instability is found to move to  $Q_z\approx 0.2 \tfrac{2\pi}{c}$.

\begin{figure}
    \centering
    \includegraphics[width=1.\linewidth]{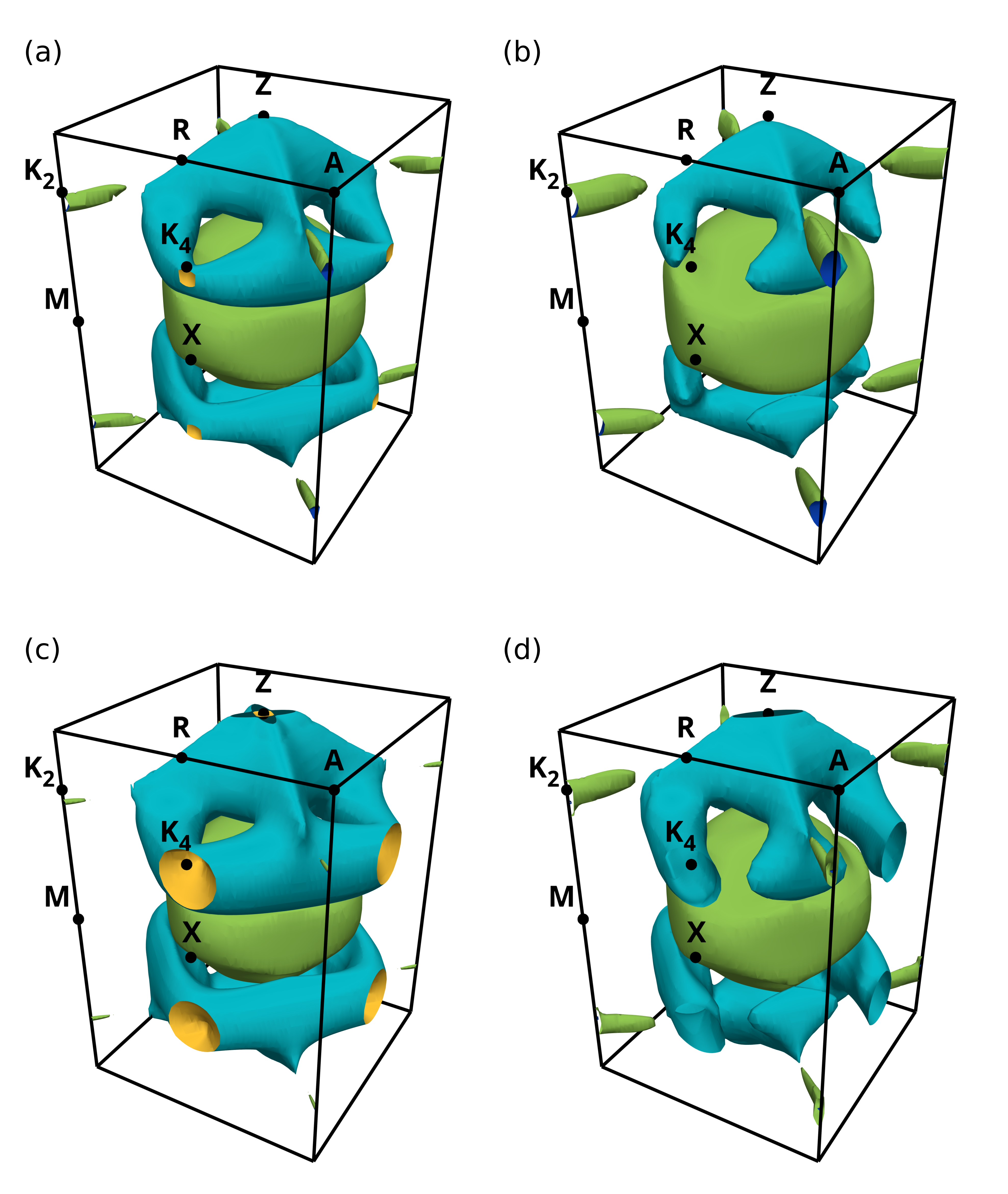}
    \caption{Fermi surfaces of non-magnetic \ruo for (a) $n = 0$, $\Delta = 0\,\mathrm{eV}$, (b) $n = 0.1$, $\Delta = 0\,\mathrm{eV}$, (c) $n = -0.1$, $\Delta = 0\,\mathrm{eV}$, and (d) $n = 0$, $\Delta = 0.4\,\mathrm{eV}$. Colors distinguish different bands.}
    \label{fig:param_fs_ruo2}
\end{figure}

The analysis using (\ref{eq:measure}) shows that the leading eigenvectors throughout the parameter space 
have $\mathcal{O} \approx 99\,\%$ overlap with the Zeeman splitting, for details see SM~\cite{SupplementRuO2}. This allows to focus on the analysis
of spin susceptibility as the response to the Zeeman splitting, which is simpler to work with than a general eigenvector. 

For example,
we can ask if the leading instability can be traced to the non-interacting spin susceptibility. 
In Fig.~\ref{fig:bubble_vs_chi_ruo2}, we compare the unfolded spin susceptibility with its corresponding non-interacting counterpart (spin bubble)
~\footnote{Note that the leading instability of $\chi$ at $Q_z=0$ corresponds to $Q_z=2\pi$ in the
unfolded susceptibility $\tilde{\chi}$, while $\tilde{\chi}(0)$ is the uniform susceptibility.}. While the non-interacting spin susceptibility
$\tilde{\chi}_0$ has quasi-degenerate maxima at $0$ and $\sim 2\pi$, the uniform susceptibility $\tilde{\chi}(0)$ is much supressed 
in the RPA solution. The multi-band character thus plays a crucial role in \ruo. The non-interacting spin susceptibility 
does not provide sufficient information about the interacting one and
the approach—where the it is simply enhanced by an effective Stoner-like interaction 
—is not applicable.
\begin{figure}
    \centering
    \includegraphics[width=0.96\linewidth]{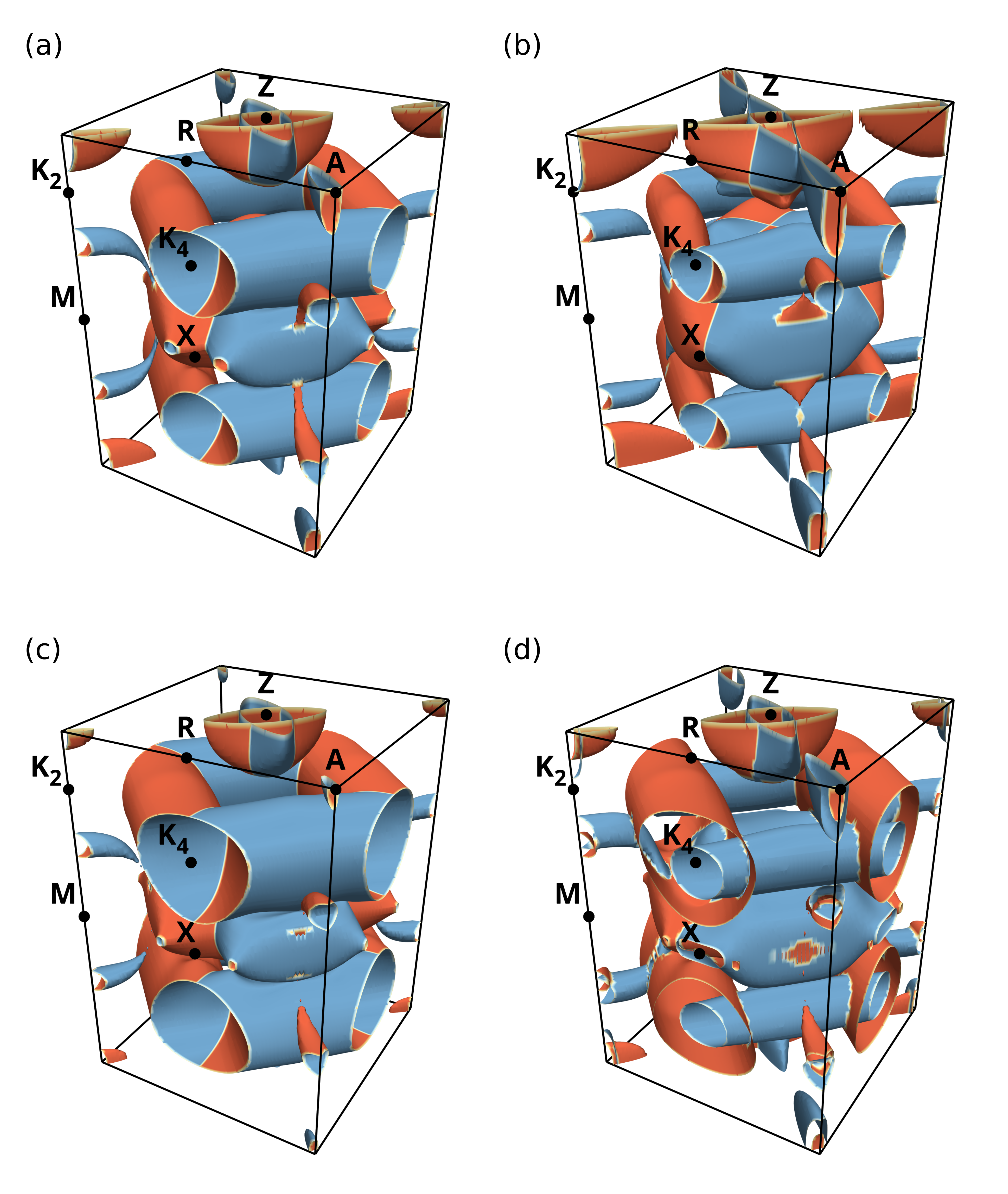}
    \caption{Fermi surfaces of antiferromagnetic \ruo for (a) $n = 0$, $\Delta = 0\,\mathrm{eV}$, (b) $n = 0.1$, $\Delta = 0\,\mathrm{eV}$, (c) $n = -0.1$, $\Delta = 0\,\mathrm{eV}$, and (d) $n = 0$, $\Delta = 0.4\,\mathrm{eV}$. Red and blue colors mark different spin polarizations, and white edges show the band-crossing.}
    \label{fig:afm_fs_ruo2}
\end{figure}

\subsection{Origin of instability for stoichiometric RuO$_{\mathbf{2}}$} 
Next, we discuss the Fermi surface geometry.
We note that the parameters used for this calculation are $U = 1.5\,\mathrm{eV}$, $\jh = 0.3\,\mathrm{eV}$, and $T = 332\,\mathrm{K}$, while $T_c = 2480\,\mathrm{K}$. For this temperature, the local magnetic moment is already saturated with size $m \approx 1 \mu_B$ per Ru site. The NM phase was obtained by a constrained paramagnetic solution. Unrealistically high $T_c$ is not surprising, since HF approximation strongly overestimates this quantity.

We note that in sections~\ref{sec:doping} and \ref{sec:stagg_pot}, the same interaction parameters are used.

\subsubsection{Fermi surface geometry}
\label{sec:FS_geometry}
The NM Fermi surface (Fig.~\ref{fig:param_fs_ruo2}~(a)) is consistent with previous theoretical studies~\cite{Berlijn2017, Zheyu2025, Peng2025}~\footnote{The deviations from \cite{Ahn2019} are due to larger Hubbard $U$ in that study.}. 
It exhibits fourfold symmetry around the $c$-axis due to the invariance of the rutile structure under the $4_2$ screw rotation. 
It consists of three closed sheets with primarily $d_{xz}$ and $d_{yz}$ character (green) and a network of tubes with $d_{x^2-y^2}$ character.

\begin{figure*}
    \centering
    \includegraphics[width=1.\linewidth]{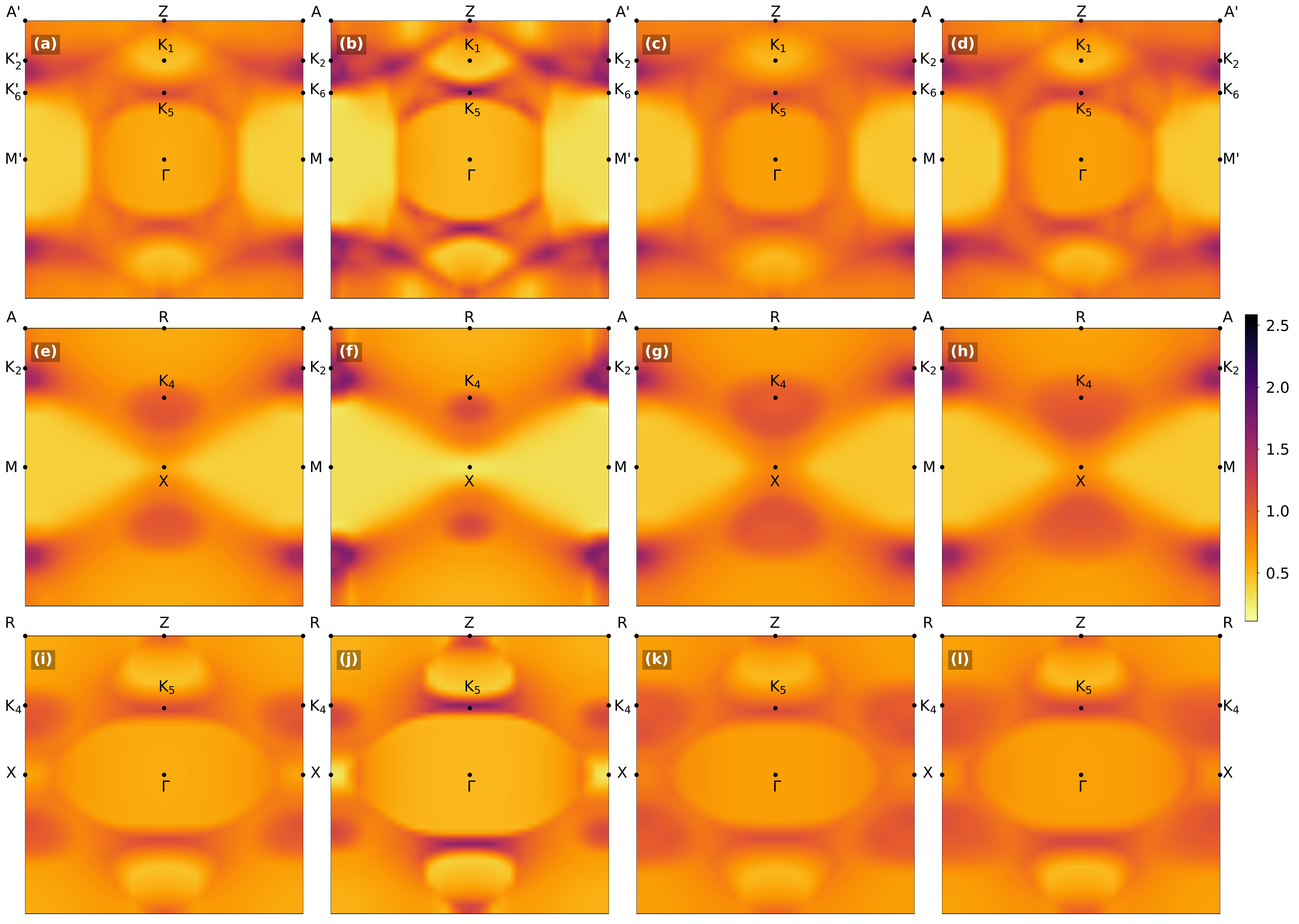}
    \caption{The cuts of $\eta(\bk)$ for (a), (e), (i) $n = 0$, $\Delta = 0\,\mathrm{eV}$; (b), (f), (j) $n = 0.1$, $\Delta = 0\,\mathrm{eV}$; (c), (g), (k) $n = -0.1$, $\Delta = 0\,\mathrm{eV}$; and (d), (h), (l) $n = 0$, $\Delta = 0.4\,\mathrm{eV}$. Each row represents different plane of cut shown by the red, blue and green plane in Fig.~\ref{fig:ruo_lattice_bz}~(b), respectively.
    The color scale is normalized to the intensity in point $K_4$ in each column.
    }
    \label{fig:eta_ruo2}
\end{figure*}

The shape of the AM Fermi surface (Fig.~\ref{fig:afm_fs_ruo2}~(a)) depends strongly on the magnitude of the ordered local moments. However, the primary feature—the spin polarization at the FS—has been reported consistently in theoretical works, independent of the computational method or interaction parameters~\cite{Ahn2019, Shao2021, Smejkal2023, Zheyu2025}.

With the onset of AM order, the tubes around $K_4$ point with $d_{x^2 - y^2}$ character become almost perfectly straight, aligned parallel
to  $[1,1,0]$ and $[1,-1,0]$ for spin-up and spin-down, respectively. The corresponding $d_{x^2-y^2}$ bands exhibit strong sublattice (Ru$_1$ and Ru$_2$) character and nearly two-dimensional dispersion due to intra-sublattice hopping (see Fig.~\ref{fig:ruo_lattice_bz}~(c)), which is modified in the NM state by hybridization with a lower-lying $d_{yz}$ band of opposite sublattice character. As the staggered order sets in the $d_{x^2 - y^2}$ and $d_{yz}$ feel an opposite Weiss field $\delta$, as indicated
in the toy model below. The majority-spin $d_{x^2 - y^2}$ band and its $d_{yz}$ partner move towards each other ending up below the Fermi level. The minority-spin $d_{x^2 - y^2}$ band and its $d_{yz}$ partner move in opposite direction and thus reducing their hybridization. The unhybibridized minority-spin $d_{x^2 - y^2}$ band then gives rise to the tubular part of the Fermi surface. The $C_4$ symmetry connecting the two Ru sites implies the 90~$\deg$ relative orientation
of the tubes with dominant weight of either sublattice.

%
\begin{equation*}
h^{\pm}_{\bk} =
\begin{pmatrix}
\epsilon - 2t\cos(k_x\pm k_y)\pm \delta & t'_{\bk}  \\
\bar{t'}_{\bk} & \mu_{\bk}\mp\delta  \\
\end{pmatrix}
.
\end{equation*}
Here $\epsilon - 2t\cos(k_x\pm k_y)$ is the band with dominant Ru$_1$ and Ru$_2$ $d_{x^2 - y^2}$ character, respectively. The hybridization $t'_{\bk}$
with the $d_{yz}$-derived band on the other sublattice is weak, $|t'_{\bk}|\ll |t|$ and the unhybridized band do not cross in the minority spin channel,
$\epsilon +2\delta -2t -\mu_{\bk}\gg |t'_{\bk}|$.
%


\subsubsection{Fermi surface hot spots}
Next, we analyze which band structure features drive the AM ordering in the weak coupling description. One may attempt to identify features with the largest contributions
to the spin susceptibility. However, the RPA spin susceptibility cannot be written as a sum of contributions from different $k$-point. While the non-interacting spin susceptibility can be decomposed in this way, the comparison of $\chi$ and $\chi_0$ above shows that their behavior, e.g., position of maxima, cannot be linked in 
a simple way. Therefore we choose a different way to access the $k$-resolved contributions to ordering.

The system orders when the free energy of AM phase becomes smaller than that of the NM phase $\langle H \rangle^{\mathrm{NM}} > \langle H \rangle^{\mathrm{AM}}$.
Within the mean-field approximation, the free energy is given by the sum of the eigenenergies of the mean-field Hamiltonian $\epsilon^{MF}_{\ell\sigma}(\bk)$ over all occupied single-particle states, plus a positive constant $C(\delta)$ proportional to the square of the Weiss field $\delta$: $\langle H \rangle_{\mathrm{MF}} = \sum_{\bk \ell\sigma} \epsilon^{MF}_{\ell\sigma}(\bk)f_{\mathrm{FD}}(\epsilon_{\ell\sigma}, T) + C(\delta)$.
The ordered phase is stable when the reduction of the first term in $\langle H \rangle_{\mathrm{MF}}^{\mathrm{AM}}$ outweighs the increase of the second term relative to the NM phase.
%
As both $\langle H \rangle^{\mathrm{NM}}$ and $\langle H \rangle^{\mathrm{AM}}$ are expressed as a sum of $k$-points we can investigate their $k$-resolved difference and
identify the main contributions
\begin{equation}
\eta(\bk) = \sum_{\ell\sigma} \left[\epsilon^{\mathrm{NM}}_{\ell\sigma}(\bk)f_{\mathrm{FD}}(\epsilon^{\mathrm{NM}}_{\ell\sigma}, T) - \epsilon^{\mathrm{AM}}_{\ell\sigma}(\bk)f_{\mathrm{FD}}(\epsilon^{\mathrm{AM}}_{\ell\sigma}, T)\right].
\end{equation}
Here, the sum runs over orbital indices (in the unit cell) and spin indices, $\ell$ and $\sigma$, respectively, $\epsilon^{\mathrm{NM}}$ denotes the energy in the NM phase, $\epsilon^{\mathrm{AM}}$ the energy in the AM phase, and $f_\mathrm{FD}(\epsilon_{\ell\sigma})$ is the Fermi-Dirac distribution. 

The function $\eta(\bk)$ has a simple interpretation. The Weiss field shifts the bands while keeping the trace over eigenvalues fixed at each $k$-point, e.g., the spin-up bands are shifted in opposite direction to the spin-down bands or an off-diagonal element pushes the band apart as in Slater antiferromagnet, which is a more relevant analogy for the present case. If the shifted partner bands are both empty of filled the net effect on $\eta(\bk)$ is zero. The main contribution thus comes from $k$-point where the 
NM partner band are both occupied (empty) while in the AM state one moves above (below) the Fermi level.

The $\eta(\bk)$ on the Brillouin zone cuts, marked in Fig.~\ref{fig:ruo_lattice_bz}~(b),
are shown in Figs.~\ref{fig:eta_ruo2}~(a),~(e),~(i). 
Three types of hot spots located near $K_2 = (0.5, 0.5, 0.35625)$,  
$K_4 = (0.5, 0, 0.25)$ 
 and $K_5 = (0, 0, 0.24)$ 
can be identified.

\begin{figure}
    \centering
    \includegraphics[width=1.\linewidth]{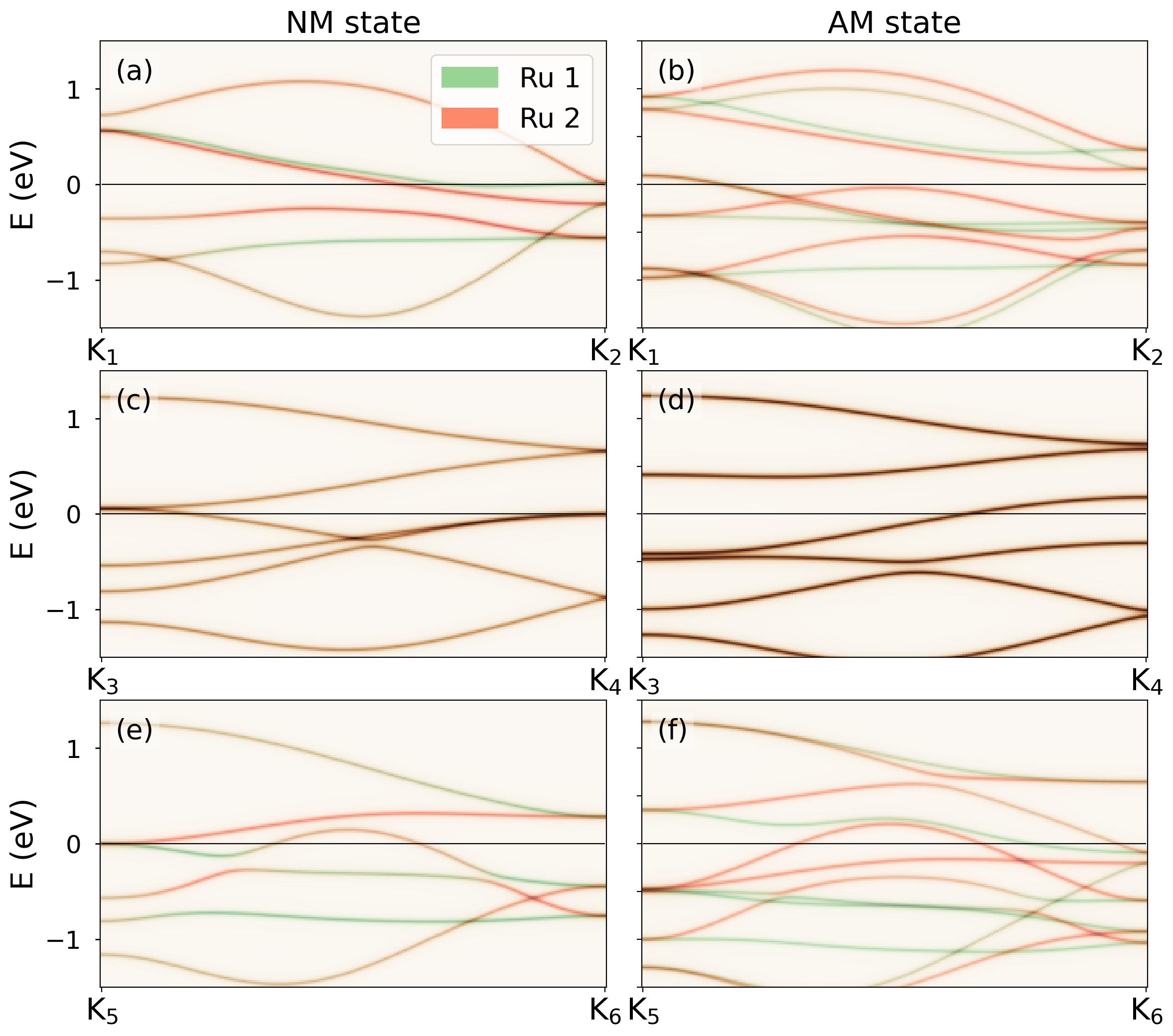}
    \caption{Site-resolved spectral function showing the band structure along the path $[K_1\text{–}K_2]$ for (a) the non-magnetic phase and (b) the altermagnetic phase. The Fermi energy is set to zero.}
    \label{fig:sf_full}
\end{figure}



The maximal value of $\eta(\bk)$ is observed near $K_2$.
The band structure along the 
$K_2$--$K_1$ line, 
approximately following  NL1 of Ref.~\cite{Sun2017}, is shown in Figs.~\ref{fig:sf_full}~(a),~(b).  
A flat band, responsible for the thin tubular sheets of the NM Fermi surface, appears near the Fermi level~\cite{Ahn2019}
and splits in the presence of magnetic order. 
The bands originating from $d_{xz}$ and $d_{yz}$ orbitals on both the same and different Ru sublattices
 are decoupled along the $MA$ line~\cite{Ahn2019}. Their dispersion is governed by intra-sublattice hopping along the $c$-axis. 
 Away from the $MA$ line, the nearest-neighbor inter-sublattice hopping start to play a role. 
 Ahn~\textit{et al.}~\cite{Ahn2019} constructed an effective model describing 
 formation of the nodal line originally identified in Ref.~\cite{Sun2017}. They showed 
 that a staggered potential gaps this nodal line in the vicinity of $K_2$ thus contributing to 
 stabilization of the AM order.
 


The second prominent hot spot of $\eta(\bk)$ is at $K_4$ on the $XR$ line. It was identified as the origin of instability in Ref.~\cite{Berlijn2017}, and observed experimentally via ARPES in Ref.~\cite{Jovic2018}. In Figs.~\ref{fig:sf_full}~(c),~(d), we show the NM and AM band structures along the $K_3$--$K_4$ line. 
The hot spot originates from the bands of $d_{x^2-y^2}$ character, discussed in Sec.~\ref{sec:FS_geometry}. The bands, which are quasi-degenerate 
on $\Gamma-X-R-Z$ planes and exactly degenerate on $X$--$R$ lines, touch the Fermi level in the vicinity of $K_4$.
The Fermi velocity on the corresponding part of the Fermi surface is significantly lower than in other parts of FS. The flatness of the bands translates to 
the large spatial extent of the hot spot.
The bands show neither site polarization along this path nor spin splitting in the AM phase, as shown in Fig.~\ref{fig:sf_full}~(d). Although the flat band splits in the AM-ordered state, the spin degeneracy is preserved in the $k_a = 0, \pi$ ($k_x$) and $k_b = 0, \pi$ ($k_y$) planes, where bands with spin-up polarization cross those with spin-down polarization. This degeneracy is protected by the glide planes $x_a = a/4$ and $x_b = a/4$, which connect the two magnetic sublattices~\cite{Ahn2019}.



The third hot spot 
lies 
around $K_5$. It is shown in Fig.~\ref{fig:eta_ruo2}~(a), and the corresponding NM and AM band structures along the $K_5$--$K_6$ line are plotted in Figs.~\ref{fig:sf_full}~(e),~(f). 
In the vicinity of $K_5$, a quadratic nodal point of $d_{xz}$ bands crosses the Fermi energy. The Fermi velocity 
near $K_5$ is higher than in the previously discussed cases. 

The band splitting in altermagnets such as \ruo differs from that in antiferromagnets. The Bloch wave functions 
in the non-magnetic state of an antiferromagnet
consist of equal superposition of the two sublattices at each k-point, i.e., no site polarization.
As a result an isolated band is not affected by the staggered Weiss field. The splitting has the form of avoided band crossings
of the back-folded bands as is well known from Slater antiferromagnets. 
Altermagnets, on the other hand, may exhibit sizable site polarization of large sections of their non-magnetic bands. 
Such portions spin-split in response to a staggered field as they are more sensitive to one sublattice than the other.
In these regions of reciprocal space the altermagnet behaves as 'crossed' ferromagnets polarized in opposite directions~\cite{Smejkal2022_2}.
In the vicinity of the hot spots we observe both band splitting mechanism the avoided band crossing as well as
the ferromagnetic-like due to site polarization.

\subsection{The effect of doping} \label{sec:doping}

\begin{figure}
    \centering
    \includegraphics[width=1.\linewidth]{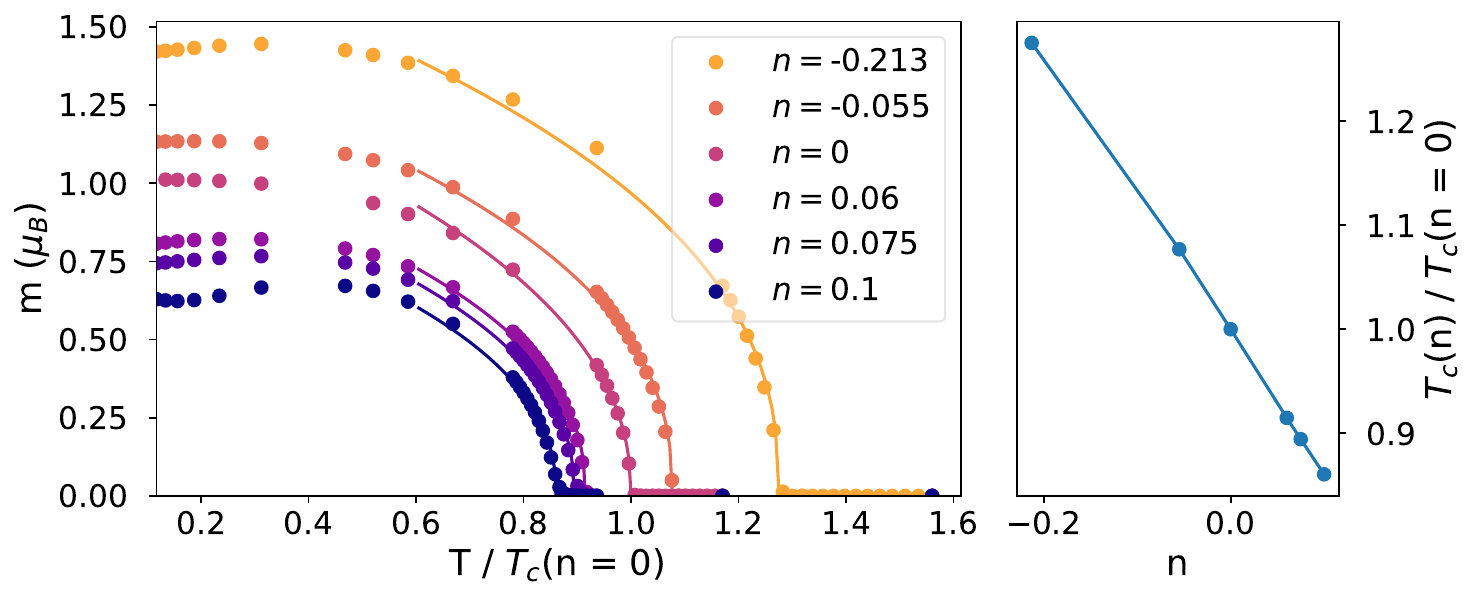}
    \caption{(letf) Temperature dependence of the local magnetic moment for various doping levels. Solid lines represent
    the $|T - T_c|^{1/2}$ fits. (right) Critical temperature of the magnetic order as a function of doping.}
    \label{fig:m_vs_doping_ruo2}
\end{figure}

The effect of doping $n$ on the temperature dependence of the local magnetic moment $m(T)$ per Ru site is shown in Fig.~\ref{fig:m_vs_doping_ruo2} (left). Negative values of $n$ correspond to hole doping and positive values to electron doping. Electron doping suppresses the magnitude of the magnetic moment, whereas hole doping enhances it. This trend is consistent with the previous study of hole doping by Smolyanyuk~\textit{et~al.}~\cite{Smolyanyuk2024}.

With hole doping, the holes predominantly enter the $d_{x^2-y^2}$ orbitals. This leads to an enhancement of the magnetic moment. The occupation numbers are shown in Fig.~5 of the Supplementary Material~\cite{SupplementRuO2}.

\begin{figure}
    \centering
    \includegraphics[width=0.9\linewidth]{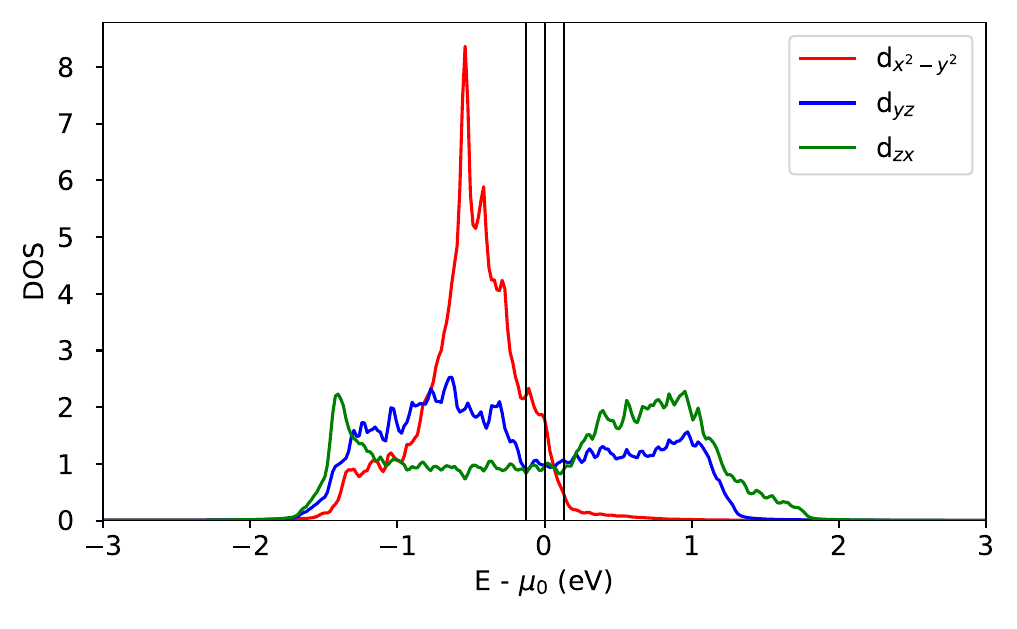}
    \caption{Orbital-resolved density of states in the NM phase of undoped \ruo. The energy is measured relative to the chemical potential of the undoped system, $\mu_0$. Vertical lines indicate the shift of the chemical potential toward negative and positive values (relative to $\mu_0$) under hole ($n=-0.1$) and electron doping ($n=0.1$), respectively.}
    \label{fig:dos_ruo2}
\end{figure}

This trend is consistent with the general expectation, as the hole doping shifts the Fermi level to higher density of states
while electron doping does the opposite, see Fig.~\ref{fig:dos_ruo2}. The effect is not uniform across the FS sheets. Mostly affected is the tubular network formed by $d_{x^2-y^2}$-derived bands. As a result the relative importance of the $K_4$ hot spot is enhanced/suppressed with hole/electron doping.

\subsection{The effect of staggered potential} \label{sec:stagg_pot}
Finally, we examine the effect of the staggered potential, which breaks the equivalence of the Ru sites. Such symmetry breaking 
may provide hints about the magnetic behavior near the surface of the sample. Our main motivation, however, is to investigate the robustness of the AM order. The stabilizing mechanism of a Slater (weakly coupled) antiferromagnet is the energy gain related to gapping of band crossings in the vicinity of the Fermi level due to a staggered Weiss field. Introducing a staggered (spin-independent) potential to the Hamiltonian opens the same gaps and thus inhibits the Slater mechanism.

\begin{figure}
    \centering
    \includegraphics[width=0.95\linewidth]{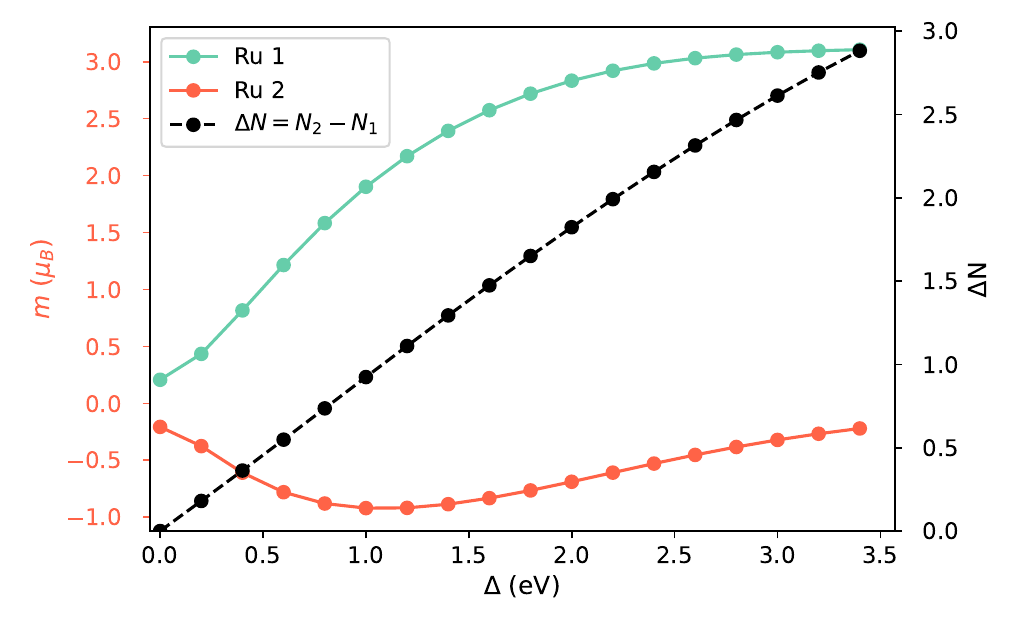}
    \caption{Site-resolved effect of the staggered potential $\Delta$ on the local magnetic moment $m$ (left axis) and the number of transferred electrons $\Delta N = N_2 - N_1$ (right axis).}
    \label{fig:delta_dep_ruo2}
\end{figure}

We have added a site-dependent potential
${H_\Delta = \tfrac{\Delta}{2} \sum_{\bR \ell \sigma} \left(c^{\dagger}_{\bR 2\ell\sigma} c^{\phantom\dagger}_{\bR 2\ell\sigma}
-  c^{\dagger}_{\bR 1\ell\sigma} c^{\phantom\dagger}_{\bR 1\ell\sigma}\right)}$
to the model. 
This potential reduces the tetragonal symmetry of both the NM and AM electronic structures.
In Fig.~\ref{fig:delta_dep_ruo2} we show the redistribution of electrons between Ru sites as well as 
the sublattice moment in the AM state, which thus becomes ferrimagnetic as there is no more symmetry connecting
the sublattices. While we performed calculations up to rather high $\Delta$, it is the small $\Delta$s that are of interest.

Contrary to na\"{\i}ve expectations based on Slater antiferromagnet picture, for example 2D square lattice model,
the staggered potential enhances rather than inhibits the magnetic order. The $\eta(\bk)$ in plot Fig.~\ref{fig:eta_ruo2}~(h),~(l)
exhibits expansion of the $K_4$ hot spot, while the other hot spots appear about the same as for $\Delta=0$. This behavior
suggests that Fermi surface nesting and related band crossing at the Fermi level is not crucial for magnetic ordering in \ruo.
Instead a ferromagnetic-like spin-dependent band splitting, which is enabled by strong site polarization of some bands
and does not rely on band crossings or proximity, plays a key role in stabilization of the magnetic order. In this sense \ruo
may be viewed as two coupled ferromagnets. The enhanced ordering tendency is related to the positive effect of hole doping on sublattice 1,
which is not compensated by the negative effect of electron doping on sublattice 2.


Finally, we note that a nonzero net magnetization and an enhancement of the magnetic moment have also been observed in epitaxial RuO$_2$/TiO$_2$ heterostructures grown on TiO$_2$ $(110)$ single-crystal substrates~\cite{Jeong2025} and in a DFT study of surface magnetism~\cite{Ho2025}. In the former case, the fourfold symmetry is broken by strain, while in the latter it is broken by the presence of a surface. Furthermore, a study of highly strained epitaxial $(110)$ \ruo films on TiO$_2$ $(110)$ substrates~\cite{Occhialini2022} reported charge redistribution between the sublattices due to changes in the bond lengths between sites. 

\section{Conclusions}
We have studied the three-orbital Hubbard model representing the Ru-$t_{2g}$ bands of \ruo using the random phase approximation.
The spin-orbit coupling was not included. An unbiased search showed that spin (i.e. orbital-diagonal) susceptibility is the dominant response. The experimentally suggested commensurate order, which gives rise to altermagnetic behavior, is found to be the leading instability
of the stoichiometric system at sufficiently low temperatures. Hole doping as well as higher temperatures lead to maxima of susceptibility
at other incommensurate wave vectors parallel to the $c$-axis. Potential spin spiral order would compete with single-ion anisotropy due to spin-orbit coupling, an option we did not investigate.

The main goal of the study was to understand the origin of incipient magnetic instability in terms of band structure and Fermi surface instability.
We have seen three 'hot spots' in the Brillouin zone, some of them identified in previous studies~\cite{Berlijn2017,Ahn2019}, which contribute significantly to stabilization of the ordered phase. We have shown that non-interacting spin susceptibility provides rather poor proxy for behavior
of the full RPA susceptibility, implying the importance of the multi-band nature of the model.



Electron doping suppresses the magnetic moment, whereas hole doping enhances the tendency of the system to order, consistent with the results of Smolyanyuk~\textit{et~al.}~\cite{Smolyanyuk2024}. The effect of moderate staggered potential as internal doping, where the positive hole doping effect of one sublattice outweighs the negative effect of electron doping on the other.


The nature of band splitting, which stabilized the magnetic order in the weak-coupling picture, is distinctly different in an altermagnet and a Slater antiferromagnet. In antiferromagnets the band splitting occurs only in the vicinity of (approximate) band crossings, while isolated bands are insensitive to staggered Weiss field,  
because each non-magnetic Bloch wave function has equal weight on the two magnetic sublattices.
In altermagnets, isolated bands can be spin-split by a staggered Weiss field if the corresponding Bloch states exhibit site polarization, 
i.e., have different weights on the magnetic sublattices.
Altermagnets can interpolate between the behavior of a Slater antiferromagnet and two weakly coupled ferromagnets or combine both as is the case studied here.

\begin{acknowledgments}

This work has received funding from the project Quantum Materials for Applications in Sustainable Technologies, Grant No. CZ.02.01.01/00/22\_008/0004572 (J.K. and D.C.), Czech Science Foundation (GA\v{C}R) project No.~GA22-28797S (D.C.) and No.~GA25-17490S (K.A.).
Computational resources were provided by Austrian Federal Ministry of Science, Research and Economy through the Vienna Scientific Cluster (VSC) Research Center and by the Ministry of Education, Youth and Sports of the Czech Republic through the e-INFRA CZ (ID:90254).
\end{acknowledgments}

\bibliography{thesis}

\end{document}


\title{Supplemental material: Incipient magnetic instability in RuO$_{\mathbf{2}}$ with random phase approximation}%

\author{Diana Csontosov\'a}
\affiliation{Department of Condensed Matter Physics, Faculty of
  Science, Masaryk University, Kotl\'a\v{r}sk\'a 2, 611 37 Brno,
  Czechia}
\author{Kyo-Hoon Ahn}
\affiliation{Institute of Physics, Czech Academy of Sciences, Cukrovarnick\'{a} 10, 162 00 Praha 6, Czechia}
\author{Jan Kune\v{s} }
\affiliation{Department of Condensed Matter Physics, Faculty of
  Science, Masaryk University, Kotl\'a\v{r}sk\'a 2, 611 37 Brno,
  Czechia}

\date{\today}%

\maketitle

\section{Methods}

\subsection{Hartree-Fock interaction term}

For the Hartree–Fock (HF) solution of the Hubbard model, we perform the mean-field decomposition of Eq.~(2) in the main text. We assume that the deviation of a single-particle operator $\hat{A}$ from its average $\langle A\rangle$ is small and therefore approximate the product of fluctuations as negligible.
\begin{equation}
    \left(\hat{A} - \langle A\rangle\right)\left(\hat{B} - \langle B\rangle\right) = \hat{A}\hat{B} - \langle B\rangle\hat{A} - \langle A\rangle\hat{B} + \langle A\rangle\langle B\rangle \approx 0.
\end{equation}
This approximation leads to the mean-field interaction Hamiltonian.
\begin{equation}
    \begin{aligned}
H_{\mathrm{HF}}
=
&\sum_{\ell}
\Big[
U
\left(
\langle c_{\ell\uparrow}^{\dagger} c_{\ell\uparrow} \rangle
c_{\ell\downarrow}^{\dagger} c_{\ell\downarrow}
+
\langle c_{\ell\downarrow}^{\dagger} c_{\ell\downarrow} \rangle
c_{\ell\uparrow}^{\dagger} c_{\ell\uparrow}
-
\langle c_{\ell\uparrow}^{\dagger} c_{\ell\uparrow} \rangle
\langle c_{\ell\downarrow}^{\dagger} c_{\ell\downarrow} \rangle
\right)
\\
&\qquad
-
U
\left(
\langle c_{\ell\uparrow}^{\dagger} c_{\ell\downarrow} \rangle
c_{\ell\downarrow}^{\dagger} c_{\ell\uparrow}
+
\langle c_{\ell\downarrow}^{\dagger} c_{\ell\uparrow} \rangle
c_{\ell\uparrow}^{\dagger} c_{\ell\downarrow}
-
\langle c_{\ell\uparrow}^{\dagger} c_{\ell\downarrow} \rangle
\langle c_{\ell\downarrow}^{\dagger} c_{\ell\uparrow} \rangle
\right)
\Big]
\\[8pt]
&+
\sum_{\substack{\ell > m \\ \sigma}}
\Big[
U'
\left(
\langle c_{\ell\sigma}^{\dagger} c_{\ell\sigma} \rangle
c_{m\bar{\sigma}}^{\dagger} c_{m\bar{\sigma}}
+
\langle c_{m\bar{\sigma}}^{\dagger} c_{m\bar{\sigma}} \rangle
c_{\ell\sigma}^{\dagger} c_{\ell\sigma}
-
\langle c_{\ell\sigma}^{\dagger} c_{\ell\sigma} \rangle
\langle c_{m\bar{\sigma}}^{\dagger} c_{m\bar{\sigma}} \rangle
\right)
\\
&\qquad
-
U'
\left(
\langle c_{\ell\sigma}^{\dagger} c_{m\bar{\sigma}} \rangle
c_{m\bar{\sigma}}^{\dagger} c_{\ell\sigma}
+
\langle c_{m\bar{\sigma}}^{\dagger} c_{\ell\sigma} \rangle
c_{\ell\sigma}^{\dagger} c_{m\bar{\sigma}}
-
\langle c_{\ell\sigma}^{\dagger} c_{m\bar{\sigma}} \rangle
\langle c_{m\bar{\sigma}}^{\dagger} c_{\ell\sigma} \rangle
\right)
\Big]
\\[8pt]
&+
\sum_{\substack{\ell > m \\ \sigma}}
\Big[ (U' - \jh)
\left( \langle c_{\ell\sigma}^{\dagger} c_{\ell\sigma} \rangle
c_{m\sigma}^{\dagger} c_{m\sigma}
+
\langle c_{m\sigma}^{\dagger} c_{m\sigma} \rangle
c_{\ell\sigma}^{\dagger} c_{\ell\sigma}
-
\langle c_{\ell\sigma}^{\dagger} c_{\ell\sigma} \rangle
\langle c_{m\sigma}^{\dagger} c_{m\sigma} \rangle 
\right)
\\
&\qquad
-
(U' - \jh)
\left( \langle c_{\ell\sigma}^{\dagger} c_{m\sigma} \rangle
c_{m\sigma}^{\dagger} c_{\ell\sigma}
+
\langle c_{m\sigma}^{\dagger} c_{\ell\sigma} \rangle
c_{\ell\sigma}^{\dagger} c_{m\sigma}
-
\langle c_{m\sigma}^{\dagger} c_{\ell\sigma} \rangle
\langle c_{\ell\sigma}^{\dagger} c_{m\sigma} \rangle 
\right)
\Big]
\\[8pt]
&+
\sum_{\substack{\ell \neq m \\ \sigma}}
\Big[
\jh
\left( \langle c_{\ell\bar{\sigma}}^{\dagger} c_{m\bar{\sigma}} \rangle
c_{m\sigma}^{\dagger} c_{\ell \sigma}
+
\langle c_{m\sigma}^{\dagger} c_{\ell \sigma} \rangle
c_{\ell\bar{\sigma}}^{\dagger} c_{m\bar{\sigma}}
-
\langle c_{\ell\bar{\sigma}}^{\dagger} c_{m\bar{\sigma}} \rangle
\langle c_{m\sigma}^{\dagger} c_{\ell\sigma} \rangle
\right)
\\
&\qquad
-
\jh
\left( \langle c_{\ell\bar{\sigma}}^{\dagger} c_{\ell\sigma} \rangle
c_{m\sigma}^{\dagger} c_{m\bar{\sigma}}
+
\langle c_{m\sigma}^{\dagger} c_{m\bar{\sigma}} \rangle
c_{\ell\bar{\sigma}}^{\dagger} c_{\ell\sigma}
-
\langle c_{\ell\bar{\sigma}}^{\dagger} c_{\ell\sigma} \rangle
\langle c_{m\sigma}^{\dagger} c_{m\bar{\sigma}} \rangle
\right)
\Big]
\\[8pt]
&-
\sum_{\substack{\ell \neq m \\ \sigma}}
\Big[
\jh
\left(\langle c_{\ell\bar{\sigma}}^{\dagger} c_{m\bar{\sigma}} \rangle
c_{\ell\sigma}^{\dagger} c_{m\sigma}
+
\langle c_{\ell\sigma}^{\dagger} c_{m\sigma} \rangle
c_{\ell\bar{\sigma}}^{\dagger} c_{m\bar{\sigma}}
-
\langle c_{\ell\bar{\sigma}}^{\dagger} c_{m\bar{\sigma}} \rangle
\langle c_{\ell\sigma}^{\dagger} c_{m\sigma} \rangle
\right)
\\
&\qquad
-
\jh
\left(\langle c_{\ell\bar{\sigma}}^{\dagger} c_{m\sigma} \rangle
c_{\ell\sigma}^{\dagger} c_{m\bar{\sigma}}
+
\langle c_{\ell\sigma}^{\dagger} c_{m\bar{\sigma}} \rangle
c_{\ell\bar{\sigma}}^{\dagger} c_{m\sigma}
-
\langle c_{\ell\bar{\sigma}}^{\dagger} c_{m\sigma} \rangle
\langle c_{\ell\sigma}^{\dagger} c_{m\bar{\sigma}} \rangle
\right)
\Big].
    \end{aligned}
\end{equation}
Indices $\ell$ and $m$ iterate over orbitals and $\bar{\sigma}=-\sigma$. In the resulting expression, the odd lines represent the Hartree terms, while the even lines correspond to the Fock terms (exchange channel). The HF self-energy $\Sigma_{\mathrm{HF}}$ is given by the matrix elements of $H_{\mathrm{HF}}$, while the terms producing a constant energy shift, $\langle\cdot\rangle\langle\cdot\rangle$, are omitted.

\subsection{Derivation of RPA bubble}

The Lehmann representation of the Green's function reads
\begin{equation}
  G_{\alpha s,\gamma s^{\prime}}(\mathbf{k}, i\nu_n) =
  \int \mathrm{d}E^{\prime}
  \frac{A_{\alpha s,\gamma s^{\prime}}(\bk, E^{\prime})}{i\nu_n - E^{\prime}}.
\end{equation}
This expression is inserted into the definition of the lattice bubble expressed on the Matsubara axis~\cite{Boehnke2011, Kunes2011}:
\begin{equation}
  \begin{aligned}
      \chi^{\alpha\beta s, \gamma \delta s^{\prime}}_0(\bq, i\nu_n, i\nu_{n'}, i \omega_m) &=
      - \frac{1}{N_k}\sum_{\bk}
      G_{\alpha s,\gamma s^{\prime}}(\bk, i\nu_n)
      G_{\delta s^{\prime}, \beta s}(\bk + \bq, i\nu_{n'} + i\omega_m)\delta_{n n'} \\
      &= - \iint
      \mathrm{d}E'
      \mathrm{d}E'' 
      \sum_{\bk}
      A_{\alpha s,\gamma s^{\prime}} (\bk, E')
      A_{\delta s^{\prime}, \beta s}(\bk + \bq, E'') \\ 
      & \times
      \dfrac{1}{i\nu_n - E'}
      \dfrac{1}{i\nu_{n} + i \omega_m - E''}\delta_{n n'}.
    \end{aligned}
  \end{equation}
Here $A_{\alpha s,\gamma s^{\prime}}$ denotes the spectral function, $i\nu_{n}$ and $i \omega_m$ represent fermionic and bosonic Matsubara frequencies, respectively, $\alpha, \beta, \gamma, \delta$ denote spin-orbital indices, and $s, s'$ are site indices. 

After summing over fermionic frequencies and using the standard expression for the Matsubara-frequency summation~\cite{Mahan}, we obtain 
\begin{equation}
  \begin{aligned}
    \label{eq:continuation}
    \chi^{\alpha\beta s, \gamma \delta s^{\prime}}_0(\bq, i \omega_m)
    &= - \frac{1}{N_k}
    \iint
    \mathrm{d}E'
    \mathrm{d}E''
    \sum_{\bk} A_{\alpha s,\gamma s^{\prime}}(\bk, E') 
    A_{\delta s^{\prime}, \beta s}(\bk + \bq, E'')
    \dfrac{f_\mathrm{FD}(E') - f_\mathrm{FD}(E'')}
    {i\omega_m + E' - E''} \\
    \chi^{\alpha\beta s, \gamma \delta s^{\prime}}_0(\bq, \omega)
    &= - \frac{1}{N_k}
    \iint
    \mathrm{d}E'
    \mathrm{d}E''
    \sum_{\bk} A_{\alpha s,\gamma s^{\prime}} (\bk, E')
    A_{\delta s^{\prime}, \beta s}(\bk + \bq, E'')
    \dfrac{f_\mathrm{FD}(E') - f_\mathrm{FD}(E'')}
    {\omega + E' - E'' + i\varepsilon} \\
    \end{aligned}
\end{equation}
where the analytic continuation is performed in the last step via the substitution $i\omega_m \rightarrow \omega + i\varepsilon$, with $\varepsilon$ infinitesimal.

By performing the substitution $\nu = E^{\prime\prime} - E^{\prime}$, renaming $E'$ to $E$, and introducing the quantity $B(\bq, \nu)$, we obtain
\begin{equation}
  \begin{aligned}
    B_{\alpha\beta s, \gamma \delta s^{\prime}}(\bq, \nu)
    &=
    \int
    \mathrm{d} E
    \sum_{\bk} A_{\alpha s,\gamma s^{\prime}} (\bk, E)
    A_{\delta s^{\prime}, \beta s}(\bk + \bq, E + \nu)
    \qty[f_\mathrm{FD}(E) - f_\mathrm{FD}(E + \nu) ],
    \label{eq:B_for_bubble}
    \end{aligned}
\end{equation}
\begin{equation}
  \begin{aligned}
    \chi^{\alpha\beta s, \gamma \delta s^{\prime}}_0(\bq, \omega)
    &=
    -  \frac{1}{N_k} \int
    \mathrm{d}\nu
    \frac{B_{\alpha\beta s, \gamma \delta s^{\prime}}(\bq, \nu)}{\omega - \nu + i\varepsilon}.
    \label{eq:chi_0_4}
    \end{aligned}
\end{equation}

Since in HF approximation the self-energy $\Sigma_{\mathrm{HF}}$ is static and represented by a matrix of real numbers, the spectral function can be written as a sum over weighted delta functions,
\begin{equation}
    A_{\alpha s,\gamma s^{\prime}}(\bk, E) = \sum_i w^i_{\alpha s,\gamma s^{\prime}}(\bk)\delta(E - E_i(\bk)),
    \label{eq:sf_delta}
\end{equation}
where $\hat{w}^i = \ket{i}\bra{i}$ is the outer product of the eigenvector of $\Tilde{H}_k = H_k + \Sigma_{\mathrm{HF}}$ corresponding to the eigenvalue $E_i$. Substituting Eq.~\eqref{eq:sf_delta} into Eq.~\eqref{eq:B_for_bubble}, we obtain
\begin{equation}
    \begin{aligned}
    B_{\alpha\beta s, \gamma \delta s^{\prime}}(\bq, \nu)
    &=
    \sum_{\bk} \sum_{ij} w^i_{\alpha s,\gamma s^{\prime}}(\bk) 
    w^j_{\delta s^{\prime}, \beta s}(\bk + \bq) \int
    \mathrm{d} E \delta(E - E_i(\bk)) \delta(E - E_j(\bk+\bq))
    \qty[f_\mathrm{FD}(E) - f_\mathrm{FD}(E + \nu) ] \\
    &= \sum_{\bk} \sum_{ij} w^i_{\alpha s,\gamma s^{\prime}}(\bk) 
    w^j_{\delta s^{\prime}, \beta s}(\bk + \bq) \delta(\nu + E_i(\bk) - E_j(\bk+\bq))
    \qty[f_\mathrm{FD}(E_i(\bk)) - f_\mathrm{FD}(E_i(\bk) + \nu) ]
    \label{eq:B_for_bubble_HF}
    \end{aligned}
\end{equation}
Finally, inserting Eq.~\eqref{eq:B_for_bubble_HF} into Eq.~\eqref{eq:chi_0_4} and applying the delta functions yields
\begin{equation}
\label{eq:bubble_final}
\begin{aligned}
\chi_0^{\alpha\beta s, \gamma\delta s^{\prime}}(\bq, \omega) = \frac{1}{N_k} &\sum_{ij} \sum_{k} w^i_{\alpha s, \gamma s^{\prime}}(\bk) w^j_{\delta s^{\prime}, \beta s}(\bk+\bq) \times \\
&\frac{f_\mathrm{FD}(E_i(\bk)) - f_\mathrm{FD}(E_j(\bk+\bq))}{\omega + i\varepsilon+ E_i(\bk) - E_j(\bk+\bq)}.
\end{aligned}
\end{equation}

\newpage
\section{Additional results}

\subsection{Temperature dependence of the instability position}

In this part, we examine the temperature dependence of instability position in the parameter regime mostly below the transition temperatures, but the NM solution was enforced. In this case, the NM phase is unstable and the susceptibility may become negative. Therefore, instead of analyzing the leading eigenvalue of $\chi(\bq)$, we focus on the smallest eigenvalues $\Lambda'(\bq)$ of the inverse static susceptibility $\chi^{-1}(\bq)$. 

\begin{figure}
    \centering
    \includegraphics[width=1.\linewidth]{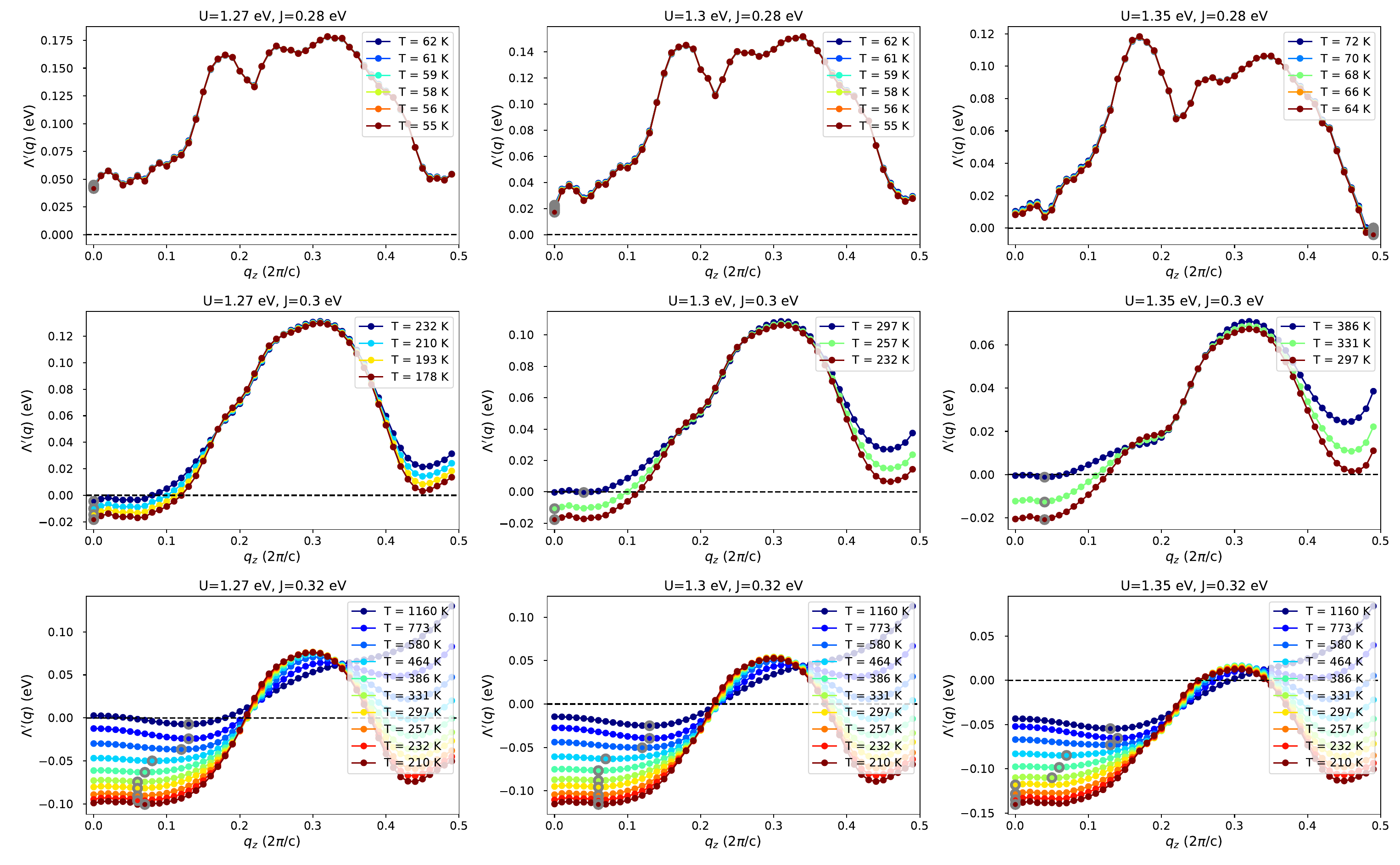}
    \caption{The temperature dependence of the smallest eigenvalues $\Lambda'(\bq)$ of the inverse static susceptibility $\chi^{-1}(\bq)$ along the $[0, 0, q_z]$ direction for various interaction parameters in the undoped system. The gray dots show the positions of instability.}
    \label{fig:temp_dep_int}
\end{figure}

\begin{figure}
    \centering
    \includegraphics[width=1.\linewidth]{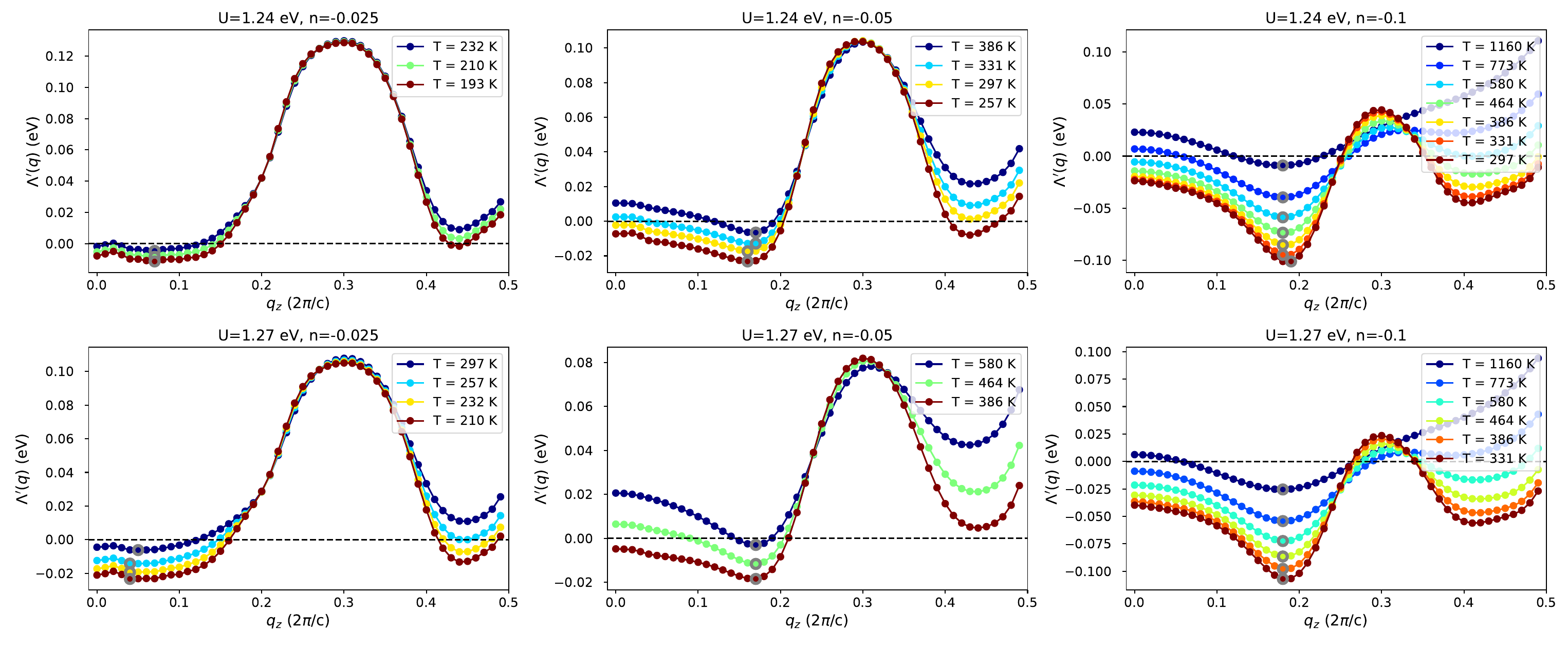}
    \caption{The temperature dependence of the smallest eigenvalues $\Lambda'(\bq)$ of the inverse static susceptibility $\chi^{-1}(\bq)$ along the $[0, 0, q_z]$ direction for various interaction parameter $U$ and hole doping $n$. The Hund's coupling is fixed at $J_{\mathrm{H}} = 0.3\,\mathrm{eV}$. The gray dots show the positions of instability.}
    \label{fig:temp_dep_doping}
\end{figure}

The temperature dependence of $\Lambda'(\bq)$ for various interaction parameters in the undoped system is shown in Fig.~\ref{fig:temp_dep_int}. 
In the cases of $J_{\mathrm{H}} = 0.3\,\mathrm{eV}$ and $J_{\mathrm{H}} = 0.32\,\mathrm{eV}$, the value of $\Lambda'(\mathbf{Q})$ shifts toward $q_z=0$ with decreasing temperature. 
For the largest $U$ and $J_{\mathrm{H}} = 0.28\,\mathrm{eV}$, the same trend toward $q_z = \frac{\pi}{c}$ is observed. This behavior is not present when hole doping is introduced (see Fig.~\ref{fig:temp_dep_doping}).

\subsection{Character of the eigenvector corresponding to the instability}

\begin{figure}
    \centering
    \includegraphics[width=1.\linewidth]{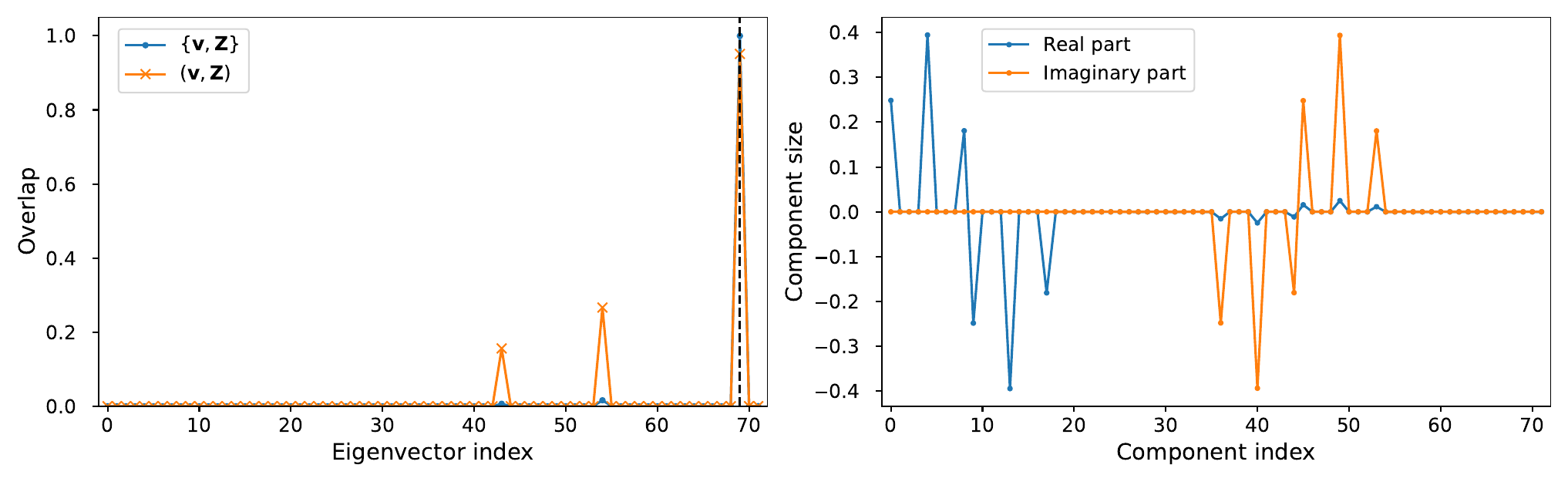}
    \caption{The absolute values of the overlaps of the eigenvectors $\mathbf{v}_i (\mathbf{Q})$ 
    with the Zeeman splitting $\mathbf{Z}$ (left). Components of the eigenvector with the largest overlap (right). The component indices of the eigenvector with dominant contribution are defined in the text. The parameters of the data point are: $U=1.35\,\mathrm{eV}, \jh=0.28\,\mathrm{eV}, n=0, T=77\,\mathrm{K}$ and corresponding order vector is $\mathbf{Q} = (0, \frac{2\pi}{a}, 0.48\frac{2\pi}{c})$.}
    \label{fig:overlaps}
\end{figure}

To determine the nature of the instability, we analyze the absolute values of the overlaps between $\mathbf{v}(\mathbf{Q})$ and $\mathbf{Z}(\mathbf{Q})$ for one point in the phase diagram shown, in the left panel of Fig.~\ref{fig:overlaps}. A vertical dashed line highlights three degenerate eigenvectors corresponding to the largest eigenvalue. 

We compare the overlaps obtained from two types of scalar products: the standard scalar product, defined as $(\mathbf{v}, \mathbf{Z}) \equiv \sum_{ij}\overline{\mathbf{v}^{ij}} \mathbf{Z}^{ij}$ (orange crosses), and the weighted scalar product $\{\mathbf{v}, \mathbf{Z}\}$ introduced in Eq.~(13) of the main text (blue dots). Since the orbital components of the eigenvectors $\mathbf{v}$ do not contribute equally, the former definition yields non-negligible overlaps with eigenvectors corresponding to different (subleading) eigenvalues. The largest overlap is $95\,\%$ and, as in the case of the latter definition, corresponds to the threefold-degenerate leading eigenvalue.

In contrast, the overlap defined in Eq.~(13) suppresses these additional contributions and clearly reveals the character of the instability. The dominant overlap (blue dots), approximately $99\,\%$, belongs to one of the three-fold degenerate leading eigenvalues, which proves the magnetic character of the instability.

The character of the dominant eigenvector is shown in the right panel of Fig.~\ref{fig:overlaps}. The eigenvector has real components on site $1$ and almost purely imaginary components on site $2$, indicating a relative rotation of the local magnetic moments on two sublattices by 90 deg. The indices of the non-negligible contributions on the $x$ axis represent the operators
\[
\begin{aligned}
\{&n_{x^2-y^2\uparrow 1}, n_{yz\uparrow 1}, n_{zx\uparrow 1},
   n_{x^2-y^2\downarrow 1}, n_{yz\downarrow 1}, n_{zx\downarrow 1}, \\
   &n_{x^2-y^2\uparrow 2}, n_{yz\uparrow 2}, n_{zx\uparrow 2},
   n_{x^2-y^2\downarrow 2}, n_{yz\downarrow 2}, n_{zx\downarrow 2}\}.
\end{aligned}
\]

\subsection{Bubble}

\begin{figure}
    \centering
    \includegraphics[width=1.\linewidth]{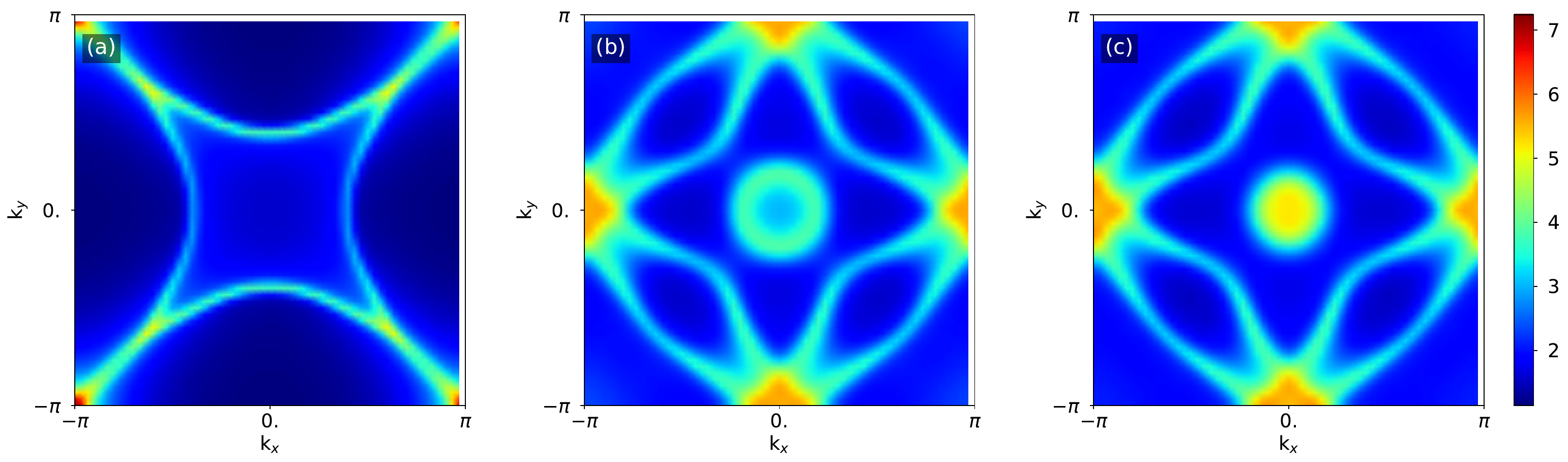}
    \caption{Momentum-resolved contribution to the leading eigenvalue of static bubble $\tilde{\chi}_0(\mathbf{Q}, \bk)$ at $\mathbf{Q} = (0, 2\pi, 0)$. The cuts are chosen to highlight the hot spots in the band structure that may be responsible for magnetic ordering at (a) $k_z = 0.35625$, (b) $k_z = 0.25$, and (c) $k_z = 0.24$ (in units of $\frac{2\pi}{c}$).}
    \label{fig:bubble_kdep_ruo2}
\end{figure}

The $k$-resolved
static spin bubble $\tilde{\chi}_{0}(\mathbf{Q}, \bk)$ at $\mathbf{Q} = (0, \frac{2\pi}{a}, 0)$ is shown in Fig.~\ref{fig:bubble_kdep_ruo2}. We plot the dependencies at $k_z = 0.35625$, $k_z = 0.25$, and $k_z = 0.24$ (in the units of $\frac{2\pi}{c}$). The maxima, with values $7.236$, $5.636$, $5.166$ and $5.715$, are located around points $K_2$, $K_4$, $K_5$ and $(0, 0.5\frac{2\pi}{a}, 0.25\frac{2\pi}{c})$, respectively. The position of the maximal contribution to $\tilde{\chi}_{0}(\mathbf{Q}, \bk)$ is in the contradiction with the one found in Ref.~\cite{Berlijn2017}.

\subsection{Occupation numbers of doped system}

\begin{figure}
    \centering
    \includegraphics[width=0.5\linewidth]{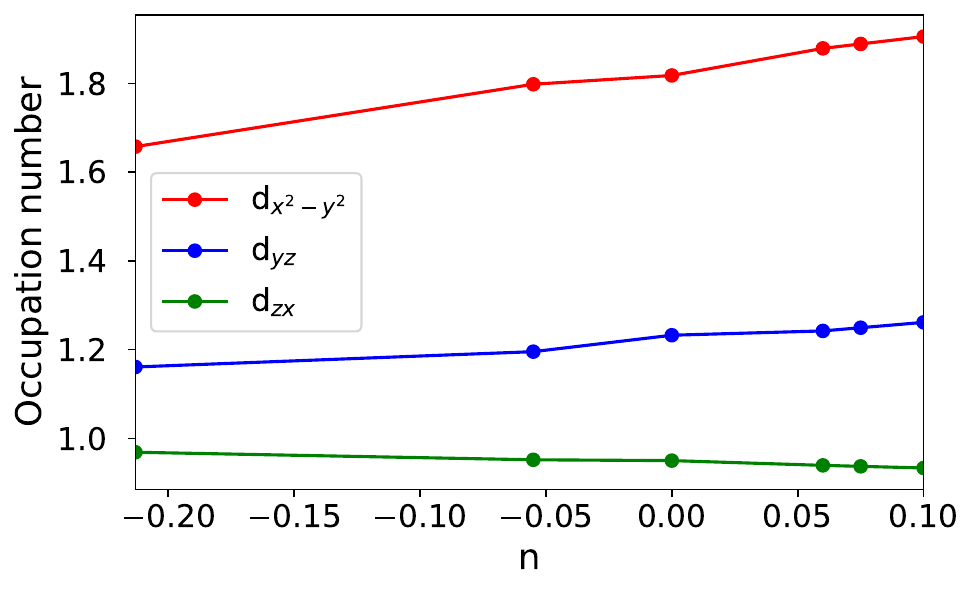}
    \caption{The occupation numbers as a function of and doping $n$ at $T=773\,\mathrm{K}$. The positive (negative) values of $n$ represent electron (hole) doping.}
    \label{fig:occ_numbers_ruo2}
\end{figure}

In this section, we show the evolution of the occupation numbers $n_{i\uparrow} + n_{i\downarrow}$ with doping and temperature, where $i$ denotes the orbital index. With electron doping, the electrons are distributed quite evenly among the orbitals. In contrast, holes predominantly occupy the $d_{x^2-y^2}$ orbitals, leading to an enhancement of the magnetic moment with hole doping.

\begin{figure}
    \centering
    \includegraphics[width=1.0\textwidth]{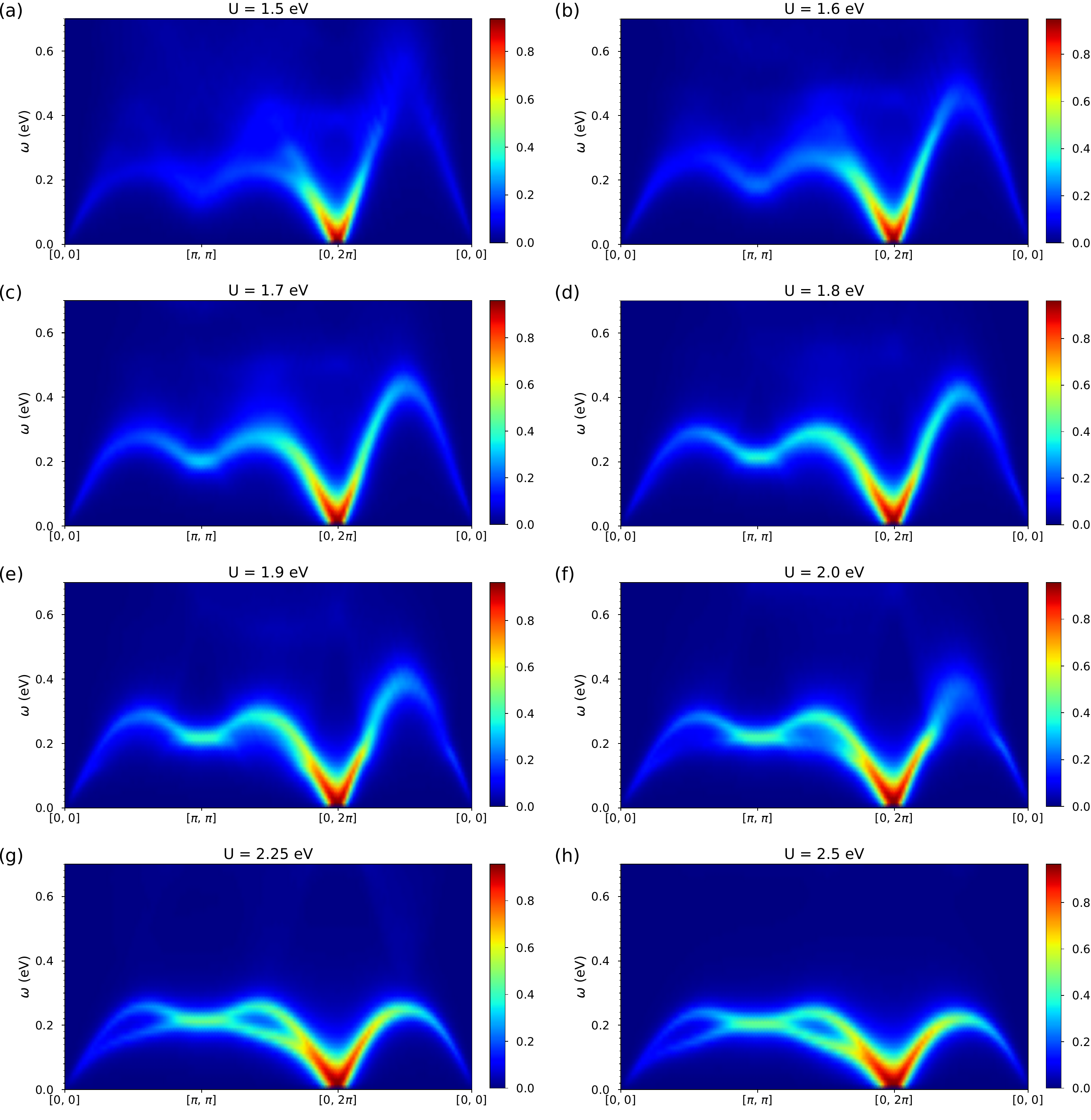}
    \caption{Imaginary part of the $\langle S_x S_x \rangle_\omega$ component of the dynamic susceptibility, showing the magnetic excitations for various values of $U$ and their corresponding magnetic moments: (a) $m = 1.0\,\mu_{\mathrm{B}}$, (b) $m = 1.24\,\mu_{\mathrm{B}}$, (c) $m = 1.41\,\mu_{\mathrm{B}}$, (d) $m = 1.51\,\mu_{\mathrm{B}}$, (e) $m = 1.57\,\mu_{\mathrm{B}}$, (f) $m = 1.61\,\mu_{\mathrm{B}}$, (g) $m = 1.70\,\mu_{\mathrm{B}}$, (h) $m = 1.75\,\mu_{\mathrm{B}}$.}
    \label{fig:sxsx}
\end{figure}

\subsection{Magnon spectra}

Finally, we show the magnetic excitations in RuO$_2$ for various values of $U$, with fixed $J_H = 0.3\,\mathrm{eV}$ and $T=297\,\mathrm{K}$. The imaginary part of the $\chi_{xx}(\omega, \mathbf{q})$ component of the dynamic susceptibility is presented in Fig.~\ref{fig:sxsx}. For all $U$ values considered, the magnon spectra exhibit a linear Goldstone mode at $\mathbf{Q} = (0, \frac{2\pi}{a}, 0)$. The unfolding procedure was applied. For clarity, we plot $\dfrac{\Im(\chi_{xx})}{\Im(\chi_{xx})+C}$, where $C=15$, to enhance the visibility of the spectral structure.



\begin{figure}
    \centering
    \includegraphics[width=1.0\textwidth]{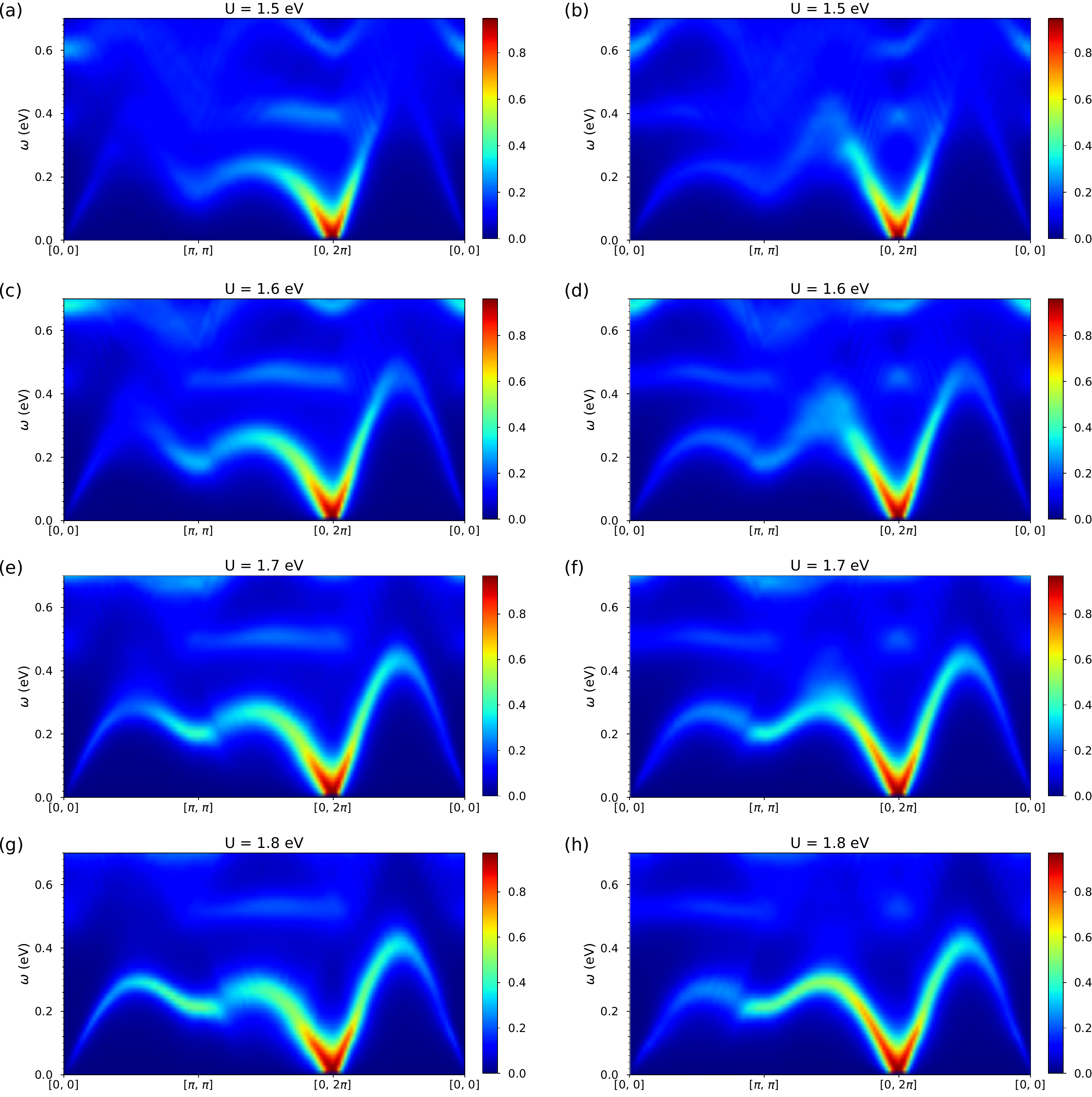}
    \caption{Imaginary part of the (a), (c), (e), (g) $\chi^{+-}$ and (b), (d), (f), (h) $\chi^{-+}$ components of the dynamic susceptibility for various values of $U$.}
    \label{fig:magnon_orb_1d5}
\end{figure}

\begin{figure}
    \centering
    \includegraphics[width=1.0\textwidth]{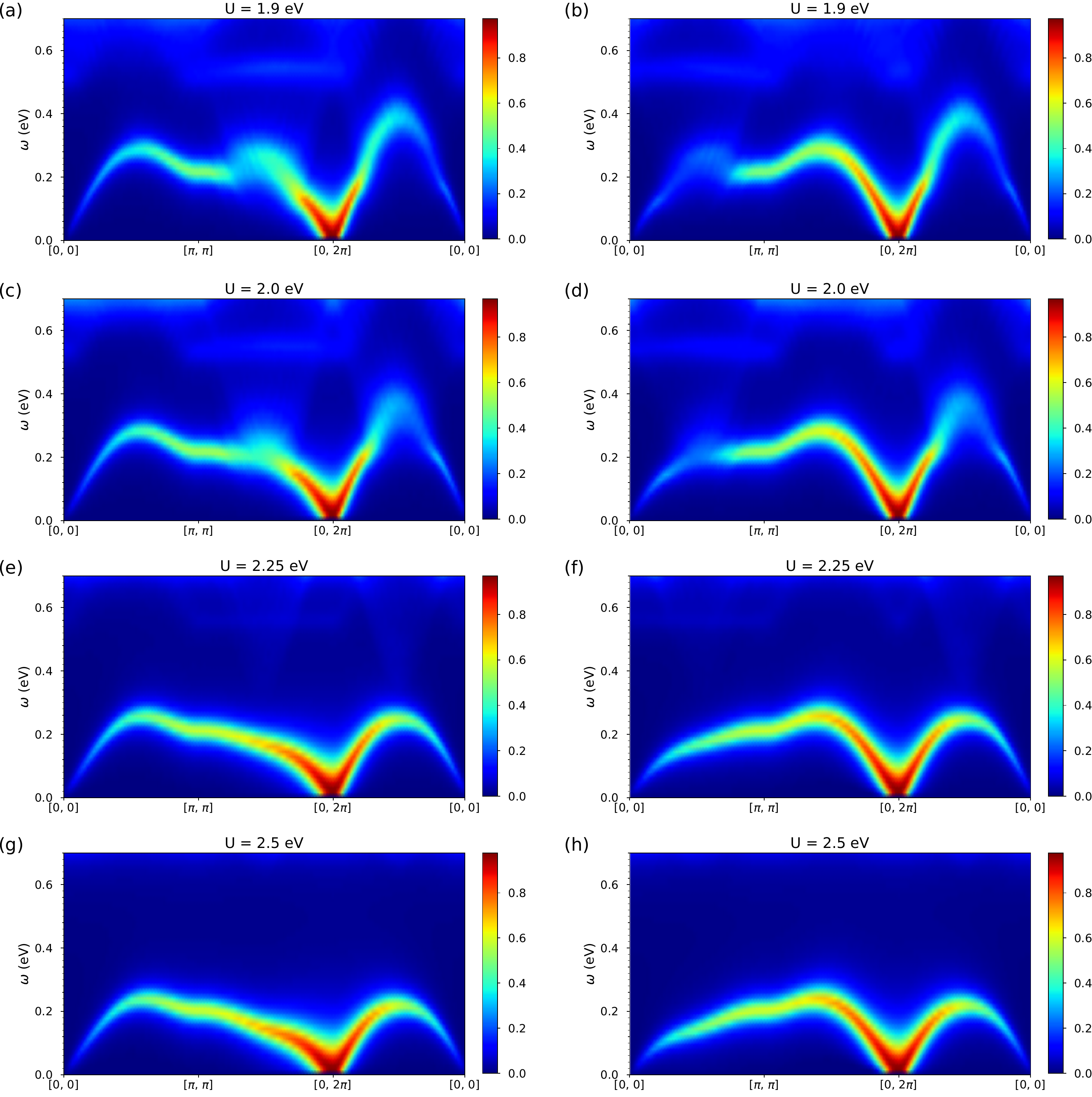}
    \caption{Imaginary part of the (a), (c), (e), (g) $\chi^{+-}$ and (b), (d), (f), (h) $\chi^{-+}$ components of the dynamic susceptibility for various values of $U$.}
    \label{fig:magnon_orb_1d9}
\end{figure}

A splitting of the magnon spectra emerges for $U > 2.0\,\mathrm{eV}$, consistent with the findings of \v{S}mejkal~\textit{et~al.}~\cite{Smejkal2023}, who employed a Heisenberg model to describe magnetic excitations. It arises from the asymmetric $\chi^{+-}$ and $\chi^{-+}$ components of the susceptibility representing the magnon branches with different chirality (see Figs.~\ref{fig:magnon_orb_1d5}~and~\ref{fig:magnon_orb_1d9}).

As $U$ decreases, the split feature vanishes, and the dispersion resembles that of a conventional antiferromagnet. As $U$ decreases, the magnon bandwidth increases and the magnon splitting becomes obscured by the overlap of the upper branch with the particle-hole continuum accompanied by pronounced damping.

\bibliography{sm_resf}